\documentclass[showpacs,floats,twocolumn,floatfix,superscriptaddress]{revtex4-1}

\bibstyle{apsrev}
\usepackage{amsmath}
\usepackage{amssymb}
\usepackage{bm}%
\usepackage{graphicx}
\usepackage{subcaption}
\usepackage{tikz}
\usepackage{color}
\usepackage{xcolor}
\usepackage[colorlinks=true,linkcolor=blue]{hyperref}%
\expandafter\ifx\csname package@font\endcsname\relax\else
 \expandafter\expandafter
 \expandafter\usepackage
 \expandafter\expandafter
 \expandafter{\csname package@font\endcsname}%
\fi
\begin{document}

\title{A lattice Boltzmann method for thin liquid film hydrodynamics}%

\author{S. Zitz}
 \affiliation{Helmholtz Institute Erlangen-N\"urnberg for Renewable Energy,\\
  Forschungszentrum J\"ulich,\\
  F\"urther Strasse 248, 90429 N\"urnberg, Germany}%
\author{A. Scagliarini}%
\email{andrea.scagliarini@cnr.it}
 \affiliation{Helmholtz Institute Erlangen-N\"urnberg for Renewable Energy,\\
  Forschungszentrum J\"ulich,\\
  F\"urther Strasse 248, 90429 N\"urnberg, Germany}%
 \affiliation{Institute for Applied Mathematics "M. Picone" (IAC), \\
Consiglio Nazionale delle Ricerche,\\
Via dei Taurini 19, 00185 Rome, Italy}%
\author{S. Maddu}
 \affiliation{Helmholtz Institute Erlangen-N\"urnberg for Renewable Energy,\\
  Forschungszentrum J\"ulich,\\
  F\"urther Strasse 248, 90429 N\"urnberg, Germany}%
  \affiliation{Center for Systems Biology, Max Planck Institute of Molecular Cell Biology and Genetics,
Pfotenhauer Strasse 108, 01307 Dresden, Germany}
\author{A.A. Darhuber}
\affiliation{Department of Applied Physics,\\
Eindhoven University of Technology, \\
PO Box 513, 5600 MB Eindhoven, The Netherlands}%
\author{J. Harting}
\email{j.harting@fz-juelich.de}
 \affiliation{Helmholtz Institute Erlangen-N\"urnberg for Renewable Energy,\\
  Forschungszentrum J\"ulich,\\
  F\"urther Strasse 248, 90429 N\"urnberg, Germany}%
\affiliation{Department of Applied Physics,\\
Eindhoven University of Technology, \\
PO Box 513, 5600 MB Eindhoven, The Netherlands}%
\date{\today}
%\revised{}

\begin{abstract}
We propose a novel approach to the numerical simulation of thin film flows,
based on the lattice Boltzmann method. We outline the basic features of the method, show in which
limits the expected thin film equations are recovered and perform validation tests.
The numerical scheme is applied to the viscous Rayleigh-Taylor instability of a
thin film and to the spreading of a sessile drop towards its equilibrium
contact angle configuration. We show that the Cox-Voinov law is satisfied, and
that the effect of a tunable slip length on the substrate is correctly
captured. We address, then, the problem of a droplet sliding on an inclined
plane, finding that the Capillary number scales linearly with the Bond number,
in agreement with experimental results. At last, we demonstrate the ability of
the method to handle heterogenous and complex systems by showcasing the
controlled dewetting of a thin film on a chemically structured substrate.
\end{abstract}

\maketitle

\newcommand{\jens}[1]{\textcolor{blue}{JENS: #1}}
\newcommand{\ts}{\textsuperscript}

\definecolor{pyblue}{HTML}{1F77B4}
\definecolor{pyorange}{HTML}{FF7F0C}
\definecolor{pygreen}{HTML}{2CA02C}
\definecolor{pyred}{HTML}{D62728}

%\tableofcontents

\section{Introduction}
Thin layers of liquids on solid surfaces are {frequently} encountered in a host
of natural and technological settings \cite{de2003capillarity,Focke}. Therefore,
understanding and controlling their stability and dynamics is a central problem
for fundamental physics, as well as for applied research in process engineering
and nanotechnology~\cite{RevModPhys.69.931,Utada}. Coating processes, for instance,
rely crucially on the mutual affinity of liquid and surface (i.e. on
wettability properties). When the liquid film is sufficiently thin, in fact, it
can become unstable, leading to the dewetting of the coated
area~\cite{RevModPhys.81.739}. From the modelling point of view, the challenge
consists in the fact that the physics of thin films is intrinsically
multiscale, for it involves phenomena ranging from the molecular scale 
at the three phase contact line, to the micro-/nano-metric size of the 
film thickness, to the size of the film as a whole, extending over the coated substrate area.

A fully resolved bottom-up atomistic
approach would be, obviously, unfeasible, if hydrodynamic regimes are to be
explored. It clearly appears that some degree of model order reduction is required. Most hydrodynamic models of thin liquid films, in the framework of the
lubrication theory, simplify the complexity of the full 3D Navier-Stokes
equations \cite{Navier,Stokes}
to one scalar transport equation (the lubrication equation) for the film thickness field
$h(x,t)$~\cite{ReynoldsLubr,RevModPhys.69.931,RevModPhys.81.1131,Mitlin}:
\begin{equation}\label{eq:lubr}
    \partial_t h = \nabla \cdot \left(Q(h)\nabla p_{\mbox{\tiny{film}}}\right)
\end{equation}
Here, $Q(h)$ is the mobility function, whose explicit form depends on the
boundary condition for the velocity at the surface (for a no-slip boundary, $Q(h) = h^3/(3 \mu)$, with
$\mu$ being the dynamic viscosity), and $p_{\mbox{\tiny{film}}}$ is the film
pressure at the free liquid surface.
Stable and reliable {direct numerical simulations} of Eq.~(\ref{eq:lubr}) require {sophisticated} numerical methods, whose {execution} is often computationally expensive~\cite{becker2003complex}.
%{Simulations of Eq.~(\ref{eq:lubr}) are numerically demanding as
%they are usually seen as a 4\ts{th} order partial differential equation. Therefore 
%not every numerical differentiation scheme is appropriate to yield stable simulations 
%further those which can provide a stable setup are usually time consuming and 
%numerically expensive.} 
Moreover, an ever-growing number of microfluidic problems requires to cope with
complex fluids rather than simple liquids, i.e. fluids with non-trivial
internal microstructure and/or complex non-Newtonian rheological
behaviour (e.g. colloidal suspensions, polymer solutions, etc.). 
The quest for an efficient multiscale numerical
method for simulating thin film hydrodynamics, versatile for the inclusion of
multiphysics features, is, thus, an ongoing endeavour. 

In this paper, we
present a novel approach to the numerical study of thin liquid films, based on
the lattice Boltzmann method (LBM) \cite{succi}. Due to the
built-in properties of the LBM, our method enjoys an outstanding
computational performance, especially on parallel architectures and graphics
processing units. 

The paper is organized as follows. We first present the numerical
model and discuss the equations of motion for the hydrodynamic fields that the model
covers. We then show that these equations effectively correspond, under certain limits,
to the lubrication equation of Reynolds.
In section~\ref{sec:results} we present
validation results including the Rayleigh-Taylor instability of thin fluid films,
the spreading of a sessile droplet on a substrate and the sliding of a
droplet on an inclined plane. After showcasing the ability of our method to
handle large and heterogeneous substrates, we present some computational
aspects including the performance of our implementation for Graphics Processing
Units (GPUs). 
An appendix is added to provide numerical tests of the validity of the
correspondence with lubrication theory (appendix~\ref{app:only}).

\section{Numerical model}\label{sec:method}

\begin{figure}
    \centering
    \includegraphics[width = 0.45\textwidth]{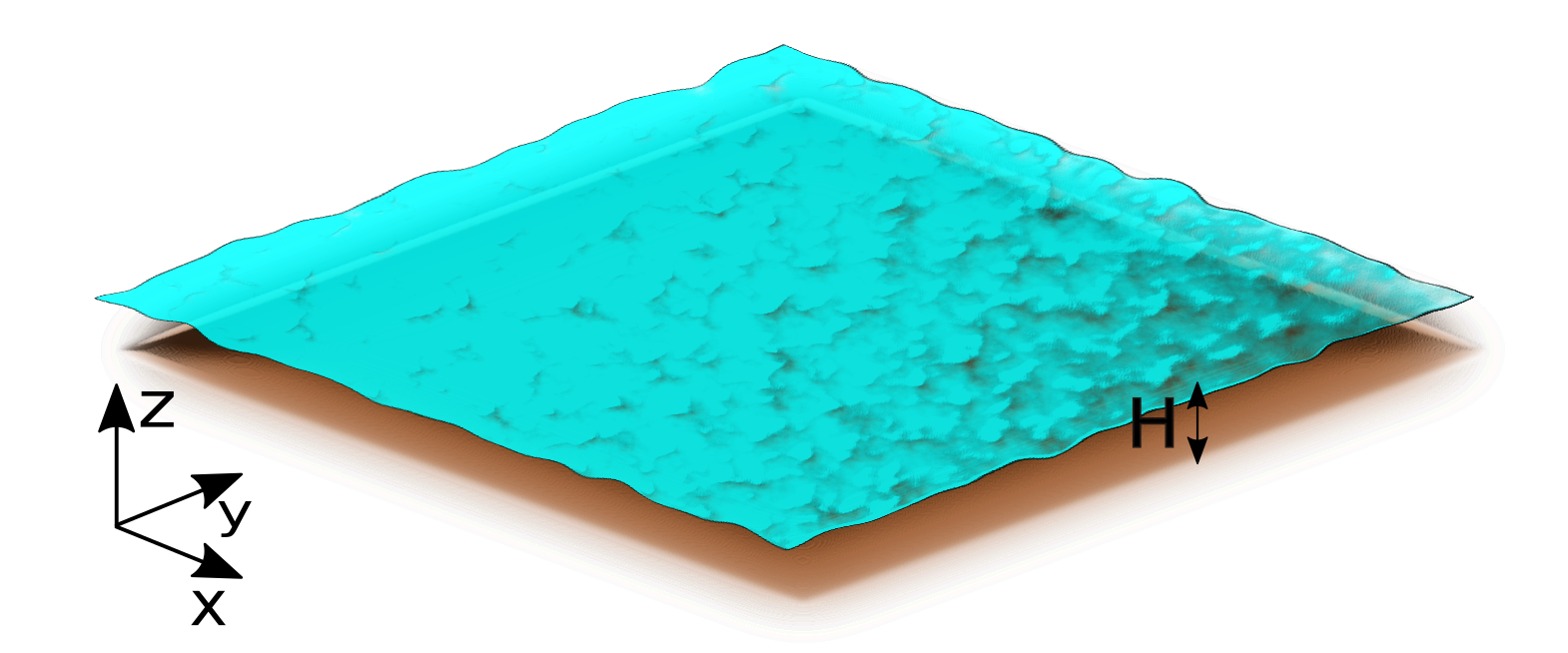}
    % Alternative there is also a very modern interpretation of the schematic figure
    % \includegraphics[width = 0.45\textwidth]{Scheme_V2.png}
    \caption{(Color online) Schematic sketch of a model system: a thin liquid film deposited on a flat substrate. The air-liquid interface is represented by the {\it height} $h(x,y,t)$. The characteristic thickness of the film is given by $H$.}
    \label{fig:scheme}
\end{figure}

When a layer of fluid is characterized by a vertical length scale $H$ much
smaller than the longitudinal one $L$, the equations of motion can be
simplified under the approximation that the ratio of the length scales, 
$\varepsilon \equiv H/L$, is small ($\varepsilon \ll1$, see Fig.~\ref{fig:scheme}).
In this limit, and for small reduced Reynolds number, $\varepsilon^2Re$
(where $Re=\frac{U L}{\nu}$, with $U$ being a characteristic velocity of the fluid system and $\nu$ being the fluid's kinematic viscosity),
the lubrication approximation tells that the dynamics is governed by equation (\ref{eq:lubr}).
Instead of directly solving Eq.~(\ref{eq:lubr}) numerically, we follow an alternative strategy.
We build our numerical model on a class of LBMs originally proposed as solvers for
the shallow water
equations~\cite{Salmon:1999:0022-2402:503,PhysRevE.65.036309,zhou2004lattice,van2010study}.
The lattice Boltzmann equation for the discrete probability density functions of a fluid system
subject to a total force (that can include both internal and external forces)
$\mathbf{F}_{\mbox{\tiny{tot}}}$,
$f_l(\mathbf{x},t)$, reads:
\begin{equation}\label{eq:LBE}
\begin{split}
&f_l(\mathbf{x}+\mathbf{c}^{(l)}\Delta t,t+\Delta t) = \\
&(1 - \omega) f_l(\mathbf{x},t) + \omega f_l^{(eq)}(\mathbf{x},t) + w_l \frac{\Delta t}{c_s^2} \mathbf{c}^{(l)} \cdot \mathbf{F}_{\mbox{\tiny{tot}}},
\end{split}
\end{equation}
where $l$ labels the lattice velocities $\mathbf{c}_l$ and runs from $0$ to $Q-1$, with $Q$ being the number
of velocities characterizing the scheme.
Algorithmically, this equation can be seen as made up of two steps. A {\it local collision}
step where the $f_l(\mathbf{x},t)$ ``relax'' towards the local equilibrium distributions
$f^{(eq)}_l(\mathbf{x},t)$ with rate  
$\omega = \Delta t/\tau$ (where $\tau$, the relaxation time, is proportional to the kinematic viscosity $\nu$): the distribution functions are substituted by their weighted average 
(with weights $\omega$ and $1-\omega$) 
with the equilibria, {with an added so-called "source" term (the last term on the right hand side of Eq. (\ref{eq:LBE})), when a force is present}. A {\it non-local streaming} step where the updated distribution functions are scattered to the nearest neighbouring sites. 
%The last term on the rhs of (\ref{eq:LBE}) includes the total force $\mathbf{F}_{\mbox{\tiny{tot}}}$ which can be in general the sum of bulk and internal (possibly space and time dependent) contributions. 
The parameters $c_s$ (the lattice speed of sound) and $w_l$ (the 
so called "weights")
depend on the geometry of the lattice and are determined under suitable constraints on 
the form of the tensorial moments in the lattice velocities up to fourth order~\cite{wolf-gladrow}.
We work with two-dimensional square lattices of side length $N\Delta x$, with lattice constant $\Delta x$ and $Q=9$.
For simplicity, we keep $\Delta t=\Delta x=1$ throughout this paper and follow the standard notation, where $c_s = 1/\sqrt{3}$ and the $\mathbf{c}^{(l)} = (c^{(l)}_x,c^{(l)}_y)$, $l=0,1,\dots,8$, are~\cite{bib:qian-dhumieres-lallemand,shan_yuan_chen_2006} 

\begin{equation}\label{eq:speeds}
\mathbf{c}^{(l)}  =
\left\{
\begin{array}{ll}
(0,0) & l = 0 \\
\left[\cos{\frac{(l-1)\pi}{4}}, \sin{\frac{(l-1)\pi}{4}} \right] &  l=1,3,5,7 \\
\sqrt{2}\left[\cos{\frac{(l-1)\pi}{4}}, \sin{\frac{(l-1)\pi}{4}} \right] & l=2,4,6,8
\end{array}
\right.,
\end{equation}
with the corresponding weights
\begin{equation}
w_l  =
\left\{
\begin{array}{ll}
\frac{4}{9} & l = 0 \\
\frac{1}{9} &  l=1,3,5,7 \\
\frac{1}{36} & l=2,4,6,8
\end{array}
\right..
\end{equation}
The equilibrium distribution functions $f_l^{(eq)}$ have to be determined to recover the
desired equations of motion for hydrodynamic fields in the long wavelength limit
(we will return to this shortly). They have, therefore, to fulfill the following 
relations involving the liquid height
\begin{equation}
    h = \sum_{l=0}^8 f_l^{(eq)},
\end{equation}
momentum
\begin{equation}
    h u_i = \sum_{l=0}^8 c_i^{(l)}f_l^{(eq)}
\end{equation}
and momentum flux tensor field
\begin{equation}\label{eq:SW_momentum_flux}
    \frac{1}{2}gh^2 \delta_{ij} + hu_i u_j = \sum_{l=0}^8 c^{(l)}_i c^{(l)}_j f_l^{(eq)},
\end{equation}
{where the left hand side coincides with the momentum flux of the shallow water equation, with the term 
$gh^2/2$ being the hydrostatic pressure in a thin fluid 
layer at rest~\cite{PhysRevE.65.036309}.}
With the usual {\it ansatz}
of a quadratic polynomial in the velocity field $\mathbf{u}$,
the equilibrium distribution functions read
\begin{equation} \label{eq:equilibria}
f_l^{(eq)}  =
\left\{
\begin{array}{ll}
h - \frac{5gh^2}{6c_s^2} - \frac{2hu^2}{3c_s^2}& l = 0 \\
\frac{gh^2}{6c_s^2} + \frac{h \mathbf{c}^{(l)}\cdot \mathbf{u}}{3 c_s^2} + \frac{h(\mathbf{c}^{(l)}\cdot \mathbf{u})^2}{2 c_s^4}-\frac{hu^2}{6c_s^2} &  l=1,3,5,7 \\
\frac{gh^2}{24c_s^2} + \frac{h \mathbf{c}^{(l)}\cdot\mathbf{u}}{12 c_s^2} + \frac{h (\mathbf{c}^{(l)}\cdot\mathbf{u})^2}{8c_s^4}-\frac{hu^2}{24c_s^2} & l=2,4,6,8
\end{array}
\right.,
\end{equation}
where $u^2 = |\mathbf{u}|^2$ is the magnitude of the velocity.
The multiscale Chapman-Enskog expansion \cite{Chapman,Enskog} of such a
LBM yields (for small ratios $Ma/Fr$
of the Mach, $Ma=u/c_s$, and Froude, $Fr = u/\sqrt{gH}$, numbers,
{corresponding also to $\sqrt{g H}/c_s \ll 1$}) the following equations
for the height and velocity fields~\cite{PhysRevE.65.036309,van2010study,Salmon:1999:0022-2402:503}
\begin{equation}\label{eq:hydro}
\begin{cases}
\begin{array}{ll}
\partial_t h + \nabla \cdot (h \mathbf{u})  = 0 & \\ 
\partial_t (h \mathbf{u}) + \nabla \cdot (h \mathbf{u}\mathbf{u}) = -gh \nabla h +\\ 
\,\,\, +  \nu \nabla^2 (h\mathbf{u}) + 2\nu \nabla (\nabla \cdot (h\mathbf{u})) +
\mathbf{F}_{\mbox{\tiny{tot}}} 
\end{array}
\end{cases},
\end{equation}
where $\nu$, the kinematic viscosity, is related to the relaxation rate $\omega$ appearing
in (\ref{eq:LBE}) via $\nu = c_s^2((2-\omega)/2\omega)\Delta t$. 
{For stability of the scheme, the condition $Fr < 1$
is also required, which is fulfilled in all our applications,
given the low values of $u$ (as discussed in more detail later on).}
Different terms contribute to the total (generalized) force\footnote{The generalized forces have
indeed the dimensions of $[\mbox{length}]^2[\mbox{time}]^{-2}$.}
$\mathbf{F}_{\mbox{\tiny{tot}}}$:
\begin{equation}\label{eq:force}
\mathbf{F}_{\mbox{\tiny{tot}}} = \mathbf{F}_{\mbox{\tiny{film}}} + \mathbf{F}_{\mbox{\tiny{fric}}} + \mathbf{F}. 
\end{equation}
In the first term the film pressure appearing in (\ref{eq:lubr}) is included as
$\mathbf{F}_{\mbox{\tiny{film}}} = -\frac{1}{\rho_0}h\nabla p_{\mbox{\tiny{film}}}$, where the film pressure $p_{\mbox{\tiny{film}}}$ 
is written as
\begin{equation}\label{eq:pfilm}
p_{\mbox{\tiny{film}}} = - \gamma (\nabla^2 h -\Pi(h))
\end{equation}
and $\rho_0$ is the (constant) liquid density 
(equal to $1$, in LBM units). 
The first term in Eq.~(\ref{eq:pfilm}) represents the capillary Laplace pressure (with $\gamma$
being the surface tension) while the second term is the disjoining pressure,
Various forms have been proposed for $\Pi(h)$ in the
literature~\cite{RevModPhys.69.931,THIELE2014399}, where here we use the expression
\begin{eqnarray}\label{eq:disjoiningP}
\Pi(h) = \kappa f(h) &=& \underbrace{(1 - \cos(\theta))\frac{(n-1)(m-1)}{(n-m)h_*}}_{\kappa}\times
\nonumber\\&&\underbrace{\left[\left(\frac{h_*}{h}\right)^n -\left(\frac{h_*}{h}\right)^m\right]}_{f(h)}.
\end{eqnarray}
In Eq.~(\ref{eq:disjoiningP}), $\theta$ is the contact angle and $h_*$
corresponds to the precursor film
thickness. The integers $n$ and $m$ are set to be $3$ and $9$, respectively. These 
are commonly chosen values in the literature~\cite{moulton_lega_2013,RevModPhys.69.931}
that correspond to a standard $6-12$ Lennard-Jones
intermolecular potential~\cite{fischer2018existence}, though other pairs
$(n,m)$ can be used (e.g.~$(2,3), (3,6), (4,10)$~\cite{1742-6596-166-1-012009,
PhysRevLett.119.204501, wedershoven2014infrared}). 
By adjusting $\kappa$ we are thus able to address the wetting properties of the substrate. 
The film pressure is specific to model thin film dynamics, in general however one can make use of other force terms e.g. to couple fluid layers which has been shown in \cite{doi:10.1002/fld.2742}. The second term on the right hand side in Eq.~(\ref{eq:force}) introduces
a friction with the substrate of the form
\begin{equation}\label{eq:forcefric}
\mathbf{F}_{\mbox{\tiny{fric}}} = -\nu \alpha_{\delta}(h)\mathbf{u}
\end{equation}
with the coefficient $\alpha_{\delta}(h)$ given by
\begin{equation}\label{eq:alphafric}
\alpha_{\delta}(h) = \frac{6h}{(2 h^2 + 6 \delta h + 3 \delta^2)}.
\end{equation}
{Here}, $\delta$ acts as a regularizing parameter, which can 
be identified with an effective slip length. 
Finally, the last term in Eq.~(\ref{eq:force}), $\mathbf{F}$ accounts for any other possible source
of forcing (e.g. the gravity component parallel to the substrate in the case of a liquid film
deposited on an inclined plate).
Equipped with such extra terms, equations (\ref{eq:hydro}) become
\begin{equation}\label{eq:hydro2}
\begin{cases}
\begin{array}{ll}
\partial_t h + \nabla \cdot (h \mathbf{u})  = 0 & \\ 
\partial_t (h \mathbf{u}) + \nabla \cdot (h \mathbf{u}\mathbf{u}) = -gh \nabla h  +\\ 
\,\, +  \nu \nabla^2 (h\mathbf{u}) \!+\! 2\nu \nabla (\nabla\! \cdot\! (h\mathbf{u}))
\!-\!\frac{1}{\rho_0}h\nabla p_{\mbox{\tiny{film}}} \!-\! \nu \alpha_{\delta}(h) \mathbf{u}\!+\! \mathbf{F}. 
\end{array}
\end{cases}
\end{equation}
Let us notice at this point that for most microfluidic applications we are actually
interested in, the advection term on the left hand side of the second equation of
(\ref{eq:hydro2}) is indeed negligible as compared to the right hand side
(the Reynolds number \cite{Reynolds,Sommerfeld} being much smaller than one).  
Analogously, the longitudinal viscous terms $\nu \nabla^2 (h\mathbf{u})$ and
$2\nu \nabla (\nabla \cdot (h\mathbf{u}))$ are of order $\varepsilon^2$ smaller 
in the ratio of length scales than the friction term $\nu \alpha(h) \mathbf{u}$
(since the former scale as $\nu H\frac{U}{L^2}$, whereas the latter as $\nu \frac{U}{H}$).
Therefore they can also be neglected.
The validity of these considerations has been numerically tested in some selected cases
(representative of typical applications) and the results are shown and discussed in the appendix.
Equations (\ref{eq:hydro2}) reduce then to
\begin{equation}\label{eq:hydro3}
\begin{cases}
\begin{array}{ll}
\partial_t h + \nabla \cdot (h \mathbf{u})  = 0 & \\ 
\partial_t (h \mathbf{u}) = -gh \nabla h -\frac{1}{\rho_0}h\nabla p_{\mbox{\tiny{film}}}
- \nu \alpha_{\delta}(h) \mathbf{u} + \mathbf{F}. 
\end{array}
\end{cases}
\end{equation}
For processes evolving on time-scales $\tilde{t}$ such that $\tilde{t}\gg \frac{h}{\alpha(h)\nu}$, one 
can consider the ``{quasi-steady}'' limit of the second of these equations 
(setting $\partial_t (h\mathbf{u}) \approx 0$), which yields
\begin{equation}\label{eq:uslaved}
\mathbf{u} \approx \frac{1}{\nu \alpha_{\delta}(h)}\left
(-gh \nabla h -\frac{1}{\rho_0}h\nabla p_{\mbox{\tiny{film}}} + \mathbf{F}\right),
\end{equation}
effectively enslaving the dynamics of $\mathbf{u}$ to that of $h$.
In the no slip limit, $\delta \rightarrow 0$, and in absence of gravity and other 
forces, Eq.~(\ref{eq:uslaved}) simplifies into
\begin{equation*}
\mathbf{u} \approx -\frac{h^2}{3 \mu}\nabla p_{\mbox{\tiny{film}}}
\end{equation*}
with the dynamic viscosity $\mu = \rho_0 \nu$. Inserting this result into the first equation of (\ref{eq:hydro3}) leads to
\begin{equation*}
\partial_t h \approx \nabla \cdot \left(\frac{h^3}{3\mu}\nabla p_{\mbox{\tiny{film}}}\right),
\end{equation*}
which is precisely the lubrication equation.
In essence, our method is, therefore, an alternative solver of the lubrication equation (at least in the inertialess regime, $Re \ll 1$, 
and for very thin films, $\varepsilon \ll 1$), that brings in, from the computational point of view, the added values of excellent scalability of the
corresponding LBM algorithm on parallel architectures, as we shall see in the
following sections. {Similar ideas have also been developed for 
reaction-diffusion equations~\cite{PhysRevA.45.5771,Kingdon_1992,CHEN1995617,WEIMAR1996207} 
and the modelling of surface tension effects by gradients of auxiliary fields~\cite{PhysRevE.54.5041, PhysRevE.67.036701} based on the color gradient method~\cite{PhysRevA.43.4320}.}

Before concluding this section, let us notice that special care has to be taken
in the implementation of the numerical scheme, when evaluating the
forcing term since it contains higher order derivatives (the gradient $p_{\mbox{\tiny{film}}}$,
which in turn includes the Laplace
 pressure $\gamma \nabla^2 h$, see Eq.(~\ref{eq:pfilm}) and, hence,
 spurious lattice effects may arise.  
We noticed, for example, that a centered scheme to calculate gradients \cite{zhou2004lattice}
does not guarantee the sufficient degree of isotropy on the lattice as, e.g., for the relaxation of
a droplet (discussed in section~\ref{sec:results}), where it led to unphysical droplet shapes.
Therefore, we use the following expressions to compute the gradients
{
\begin{equation}\label{eq:derivative}
    \nabla \phi(\mathbf{x}) = 3\sum_{l=0}^8w_l\mathbf{c}^{(l)} \phi(\mathbf{x}+\mathbf{c}^{(l)})+O(\nabla^3),
\end{equation}
and the Laplacian 
\begin{align}\label{eq:n_laplace}
    \nabla^2 \phi(\mathbf{x}) =& \frac{1}{6}\bigg[4\sum_{l = {\rm odd}} w_l \phi(\mathbf{x}+\mathbf{c}^{(l)})\nonumber\\ 
    &+ 1\sum_{l = {\rm even}} w_l \phi(\mathbf{x}+\mathbf{c}^{(l)}) - 20\phi(\mathbf{x}) \bigg] + O(\nabla^4),
\end{align}
respectively \cite{doi:10.1137/S1064827599357188,THAMPI20131}}, for a generic scalar field $\phi$ 
(be it the height field $h$, the pressure $p_{\mbox{\tiny{film}}}$ or a position dependent surface tension field).
Besides the higher degree of the isotropy, the scheme 
(\ref{eq:derivative}-\ref{eq:n_laplace}) has the advantage of employing directly the set of lattice Boltzmann speeds. 

\section{Results}\label{sec:results}
{Below we present results from numerical simulations using the method introduced in the previous section. For all simulations, we apply periodic boundary conditions in the $X$-$Y$ plane.}
\subsection{The Rayleigh-Taylor instability}

\begin{figure*}

  \begin{subfigure}[t]{0.24\textwidth}
    \centering\includegraphics[width=1.0\textwidth]{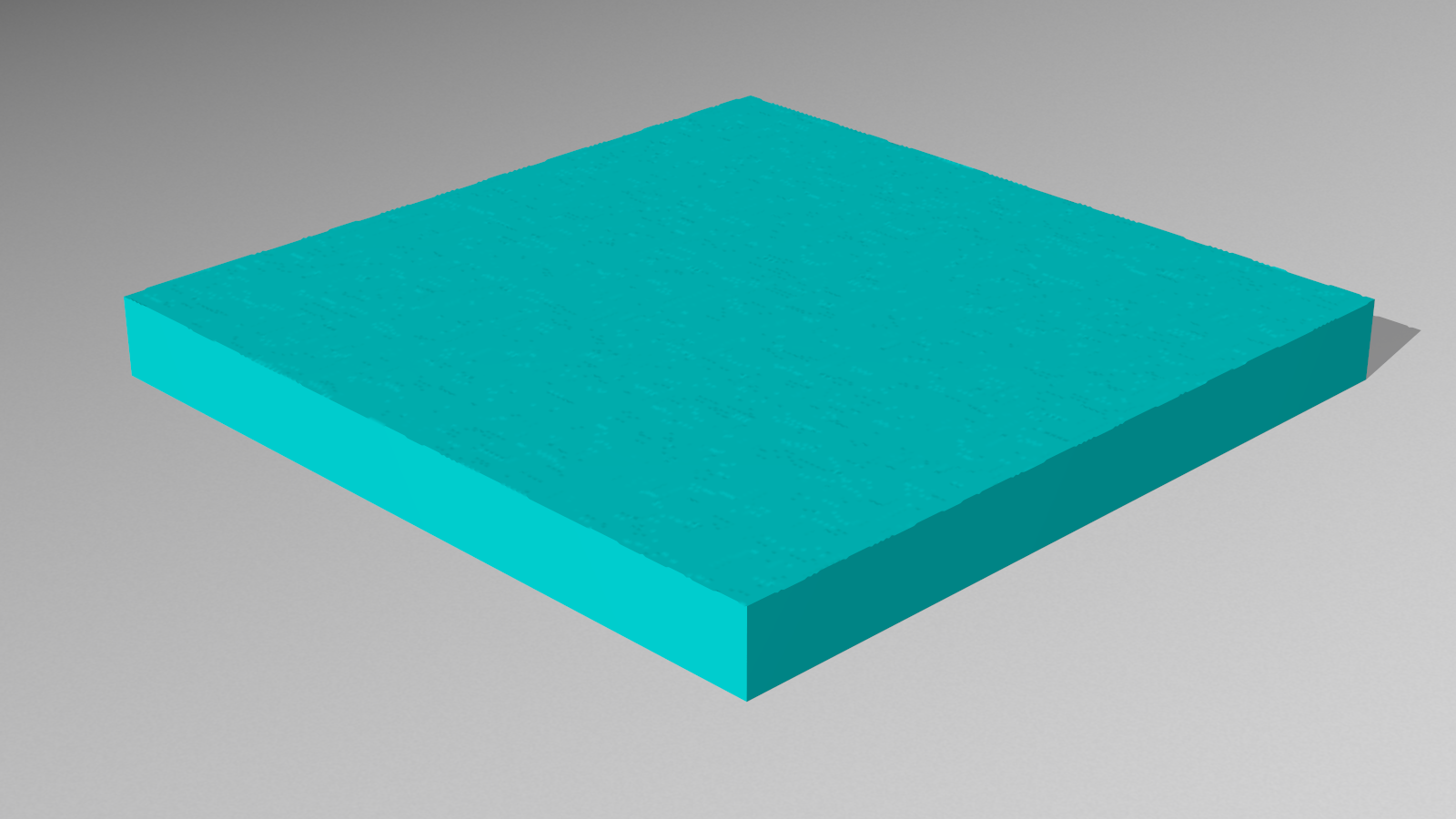}
    \caption{t $=9000\Delta$t}
  \end{subfigure}
  \begin{subfigure}[t]{0.24\textwidth}
    \centering\includegraphics[width=1.0\textwidth]{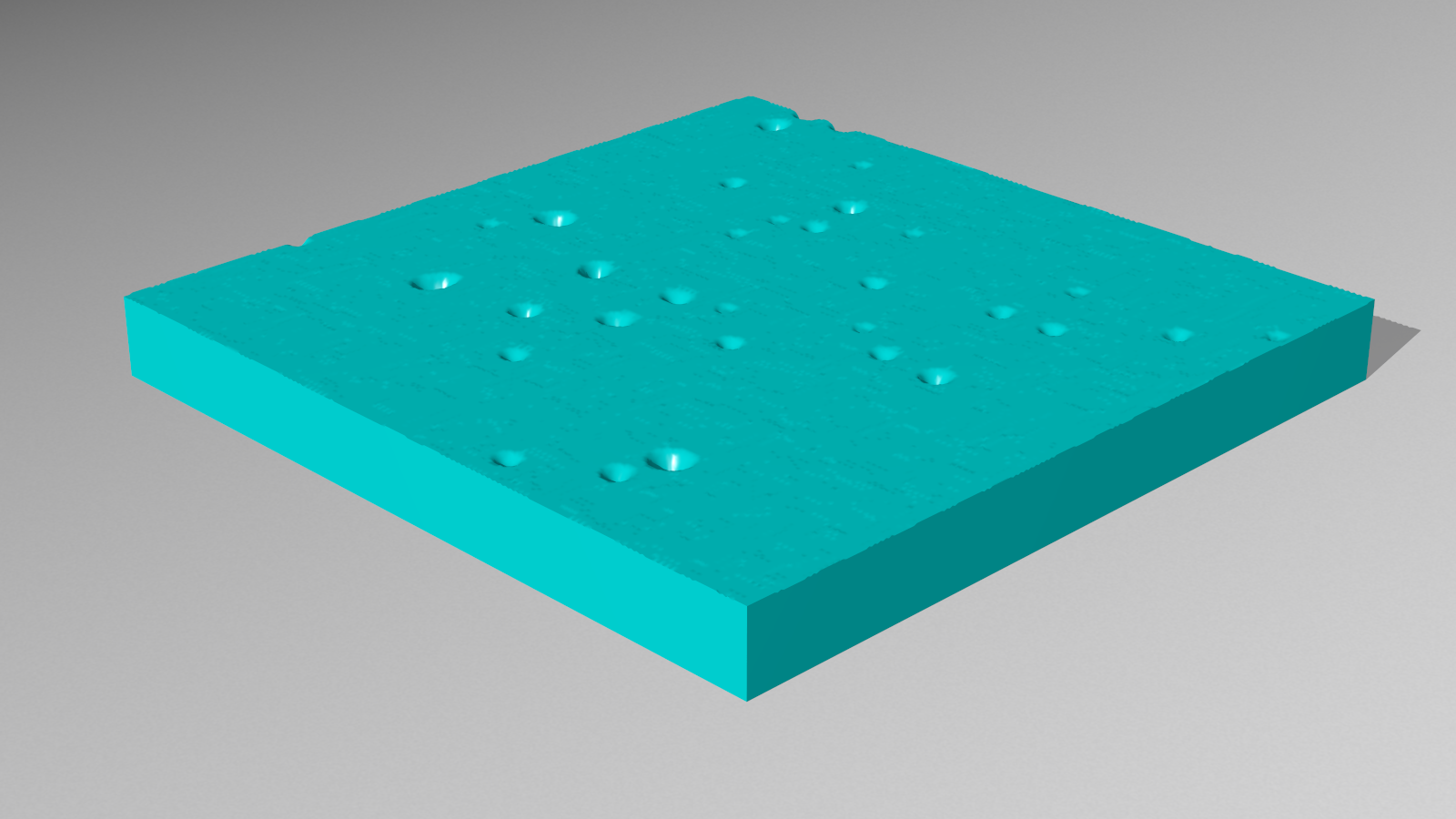}
    \caption{t $=14000\Delta$t}
  \end{subfigure}
  \begin{subfigure}[t]{0.24\textwidth}
    \centering\includegraphics[width=1.0\textwidth]{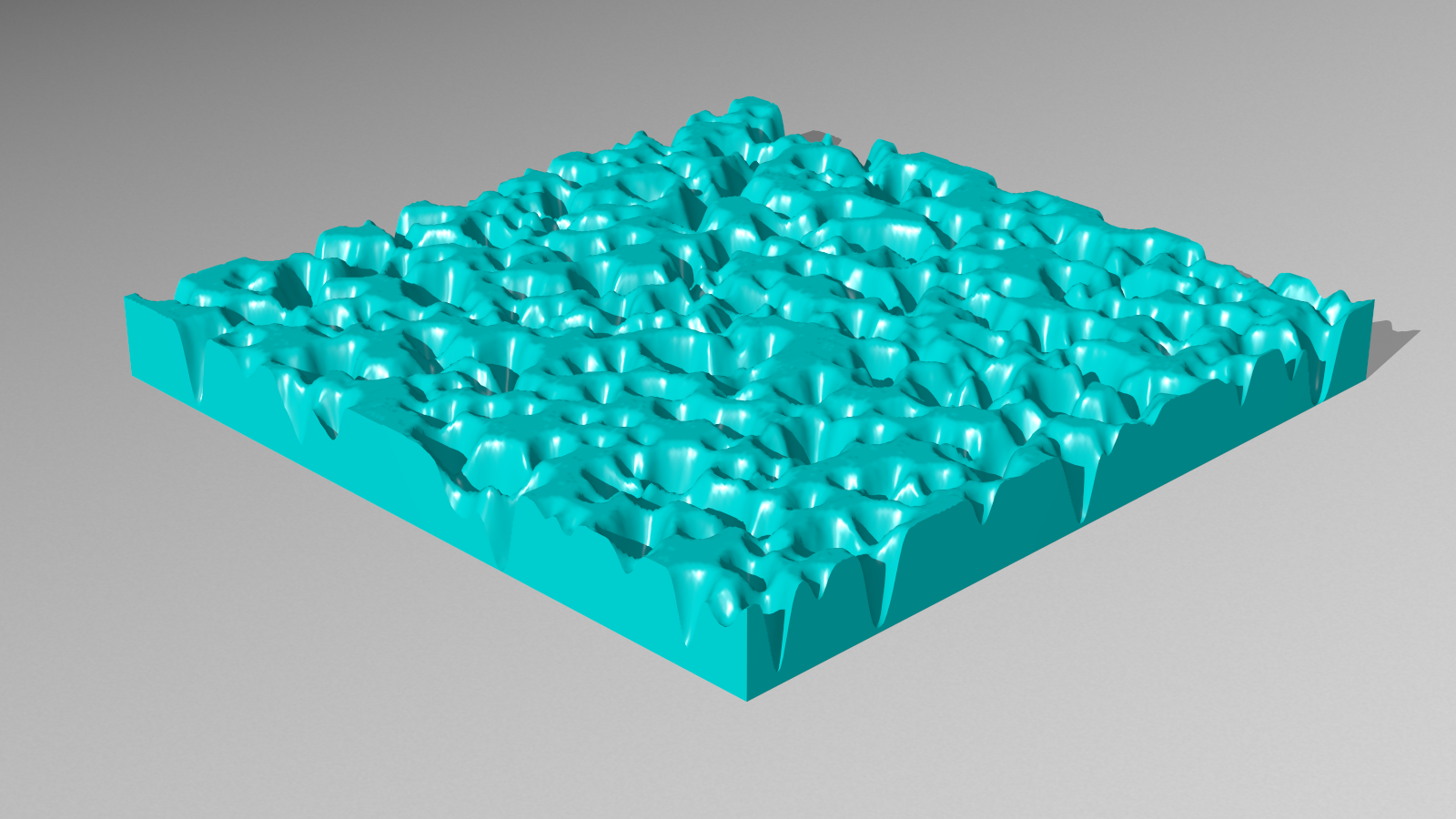}
    \caption{t $=19000\Delta$t}
  \end{subfigure}
  \begin{subfigure}[t]{0.24\textwidth}
    \centering\includegraphics[width=1.0\textwidth]{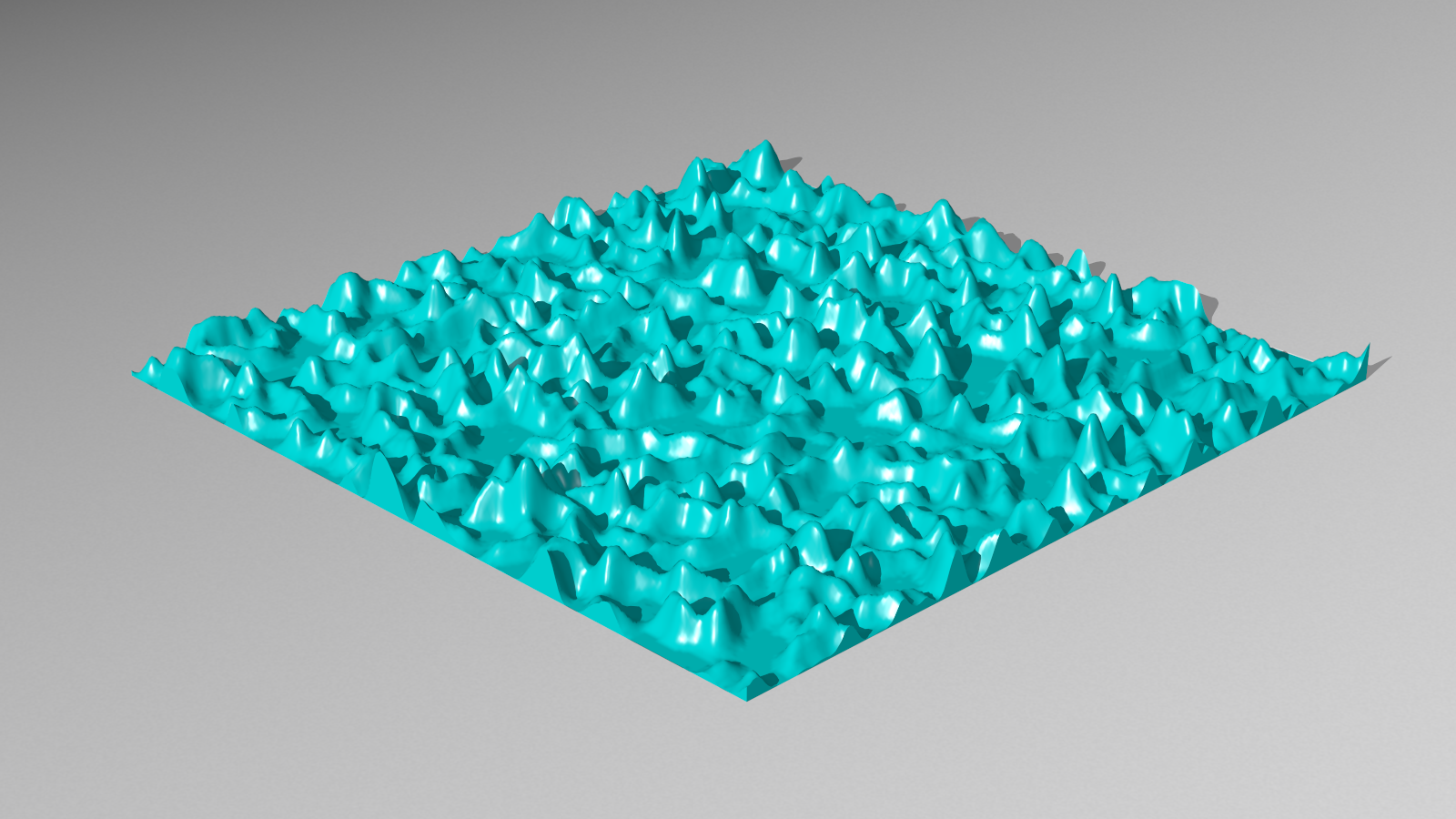}
    \caption{t $=35000\Delta$t}
  \end{subfigure}
  \caption{(Color online) Time evolution of the free surface for the Rayleigh-Taylor instability at $\tau_{\mbox{\tiny{cap}}}\approx50, 75, 100, 188$. For a more clear visualization we only show a small patch of size $256\times256$ centered in the middle of the $2048\times 2048$ domain. The fluctuations of earlier states still follow the linear stability analysis (See Fig.~\ref{fig:RTI} for the {the power
spectrum  of the  height fluctuations versus wavenumber}).}
  \label{fig:RTI_evolution}
\end{figure*}

\begin{figure}
    \centering
    \includegraphics[width=0.45\textwidth]{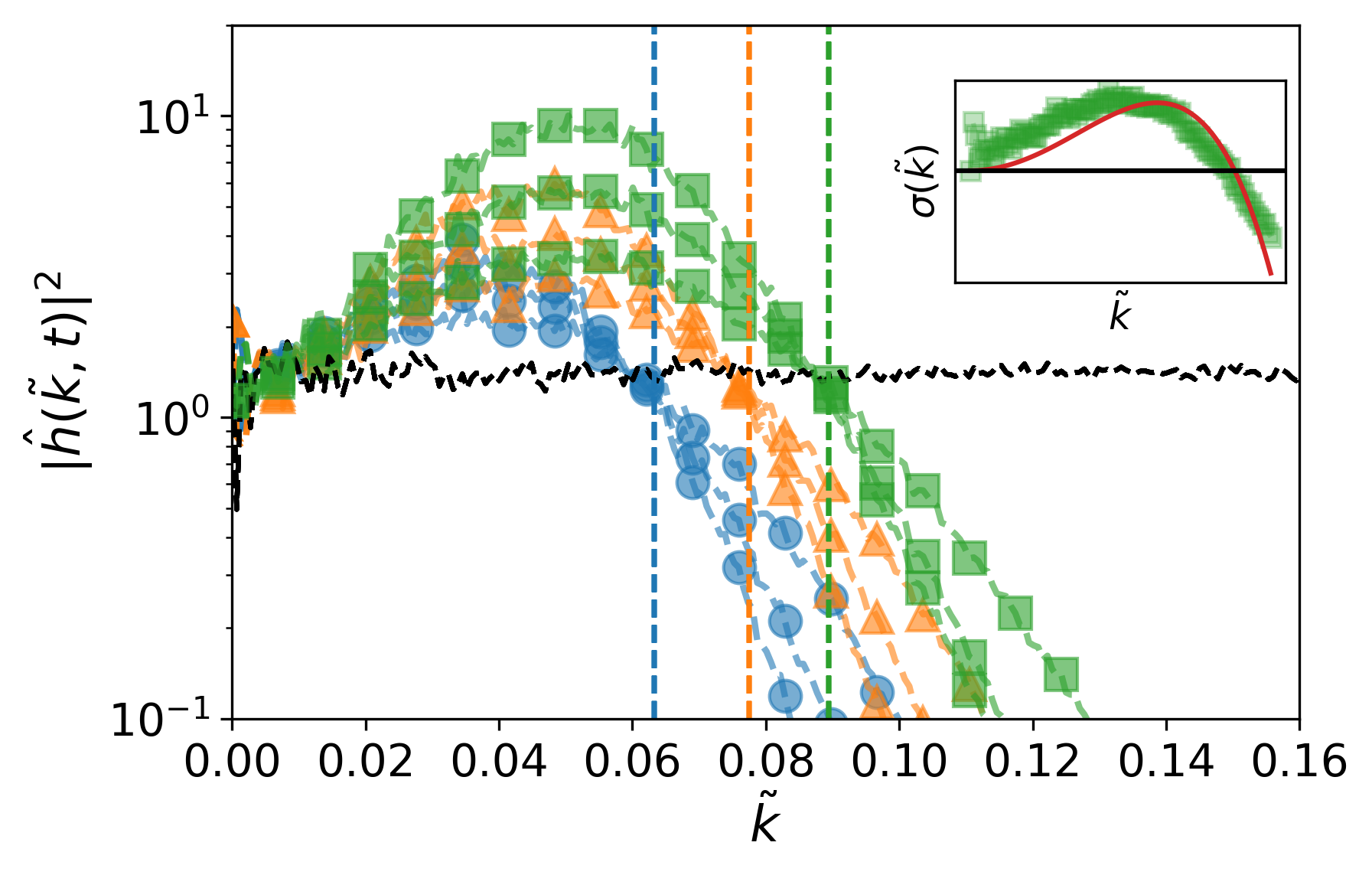}
    \caption{(Color online) {Power spectrum of the height fluctuations versus wavenumber.} The different colors and symbols belong to different values of graviational acceleration, $g=4\cdot 10^{-5}$ is given by blue circles (\textcolor{pyblue}{$\bullet$}), $g=6\cdot 10^{-5}$ by orange triangles (\textcolor{pyorange}{$\blacktriangle$}) and $g=8\cdot 10^{-5}$ is given by green squares (\textcolor{pygreen}{$\blacksquare$}). Same colored lines are taken at different time steps. In the inset we show the growth rate $\sigma(k)$ for the largest value of $g$ (symbols) and the theoretical growth rate according to Eq.~(\ref{eq:RTgrowth}) (solid line).}
    \label{fig:RTI}
\end{figure}

The Rayleigh-Taylor instability occurs when a denser fluid is accelerated 
against a less dense one~\cite{Rayleigh,Taylor,KULL1991197,Sharp}. This can be the case, for instance, 
for a liquid film coating a ceiling, under the action of gravity. In such a configuration
gravity tends, of course, to deform (and eventually disrupt) the film, while surface tension has a stabilizing effect.
As a result of these competing mechanisms, any surface perturbation is stable or unstable
depending on whether its characteristic wavenumber $k$ is smaller or larger than a certain
critical value $k_c$. Linear stability analysis calculations in the framework of lubrication theory provide the following growth rate $\sigma(k)$:
\begin{equation}\label{eq:RTgrowth}
    \sigma(k) = \frac{\rho g h_0^3}{3\mu}(k^2 - l_{\mbox{\tiny{cap}}}^2 k^4),
\end{equation}
where $l_{\mbox{\tiny{cap}}} = (\gamma/g)^{1/2}$ is the capillary length. 
Unstable (stable) modes correspond to
$\sigma(k)>0$ ($\sigma(k)<0$) and the critical wavenumber is, therefore, such that $\sigma(k_c)=0$, i.e. 
$k_c = 1/l_{\mbox{\tiny{cap}}}$.
On a lattice of size $2048 \times 2048$ nodes, we initialize the film height according to
\begin{equation}
 h(\mathbf{x},0) = h_0(1 + \varepsilon(\mathbf{x})),
\end{equation}
with $\varepsilon$ a random variable homogeneously distributed in $[1\cdot 10^{-4},-1\cdot 10^{-4}]$ and $h_0 = 1$. Forcing should always be below a certain threshold. Thus, for the gravitational acceleration we choose values within the interval $|g| = [4,8]\cdot 10^{-5}$. Furthermore, we fix the surface tension to be $\gamma=0.01$. This results in critical wavenumbers ranging from $k_c= 0.06$ to $k_c = 0.09$.
Fig.~\ref{fig:RTI_evolution} shows snapshots of the free surface from various time steps, where the growth of the perturbations is shown as time increases. The last panel is already beyond the linear regime.

We consider the time evolution of the power spectrum of the height field fluctuations 
(around the mean), defined as 
\begin{equation}\label{eq:powerspec}
    E(k,t) = \oint_{\Omega_k} |\hat{\delta h}(\mathbf{k},t)|^2 d\Omega_k,
\end{equation}
where
\begin{equation}\label{eq:spectra}
    \hat{\delta h}(\mathbf{k},t) = \int e^{-2\pi i\mathbf{k}\cdot\mathbf{x}} (h(\mathbf{x},t)-h_0)~\mathrm{d}\mathbf{k},
\end{equation}
with $\mathbf{k}=(k_x,k_y)$. $\Omega_k$ denotes the circle in $\mathbf{k}$-space (i.e. 
$\Omega_k = \{(k_x, k_y) | k_x^2 + k_y^2 = k^2 \}$). {Since we work in a discretized
system we have to smear out the circle $\Omega_k$ with some small $\delta k$. Therefore, strictly speaking 
the integral is not computed around the circle $\Omega_k$ but around some small annulus 
$\Omega_{k + \delta k}$.}
The spectra are shown in Fig.~\ref{fig:RTI}. The various colors and symbols of Fig.~\ref{fig:RTI} relate to different values of gravitational acceleration. For every set we consider the spectra at three equally scaled 
times $\tilde{t}=t/\tau_{\mbox{\tiny{cap}}}$, where
\begin{equation}
    \tau_{\mbox{\tiny{cap}}} = \frac{\mu l_c}{\gamma},
\end{equation}
$\tilde{t}=50, 75, 100$. The values of $k_c$ correspond to the points where the colored lines with symbols cut the black dashed line. The horizontal colored dashed lines mark the theoretical values for $k_c$. We observe good agreement of theoretical and numerical values. In the inset of Fig.~\ref{fig:RTI} we plot the growth rate for the data of $g = 8\cdot 10^{-5}$ together with the theoretical expression given by Eq.~(\ref{eq:RTgrowth}) (solid line).

Consistently with the random initialization, at $\tilde{t}=0$ the spectrum is a constant (black dashed line). 
For $\tilde{t}>0$, $E(k,t)$ develops a profile that grows in time for $k<k_c$, while
it is damped out for $k>k_c$, in agreement with the expectation from the theory.

\subsection{A spreading droplet}
\begin{figure*}
    \centering
    \begin{subfigure}[t]{0.45\textwidth}
        \centering
        \includegraphics[width=1.0\textwidth]{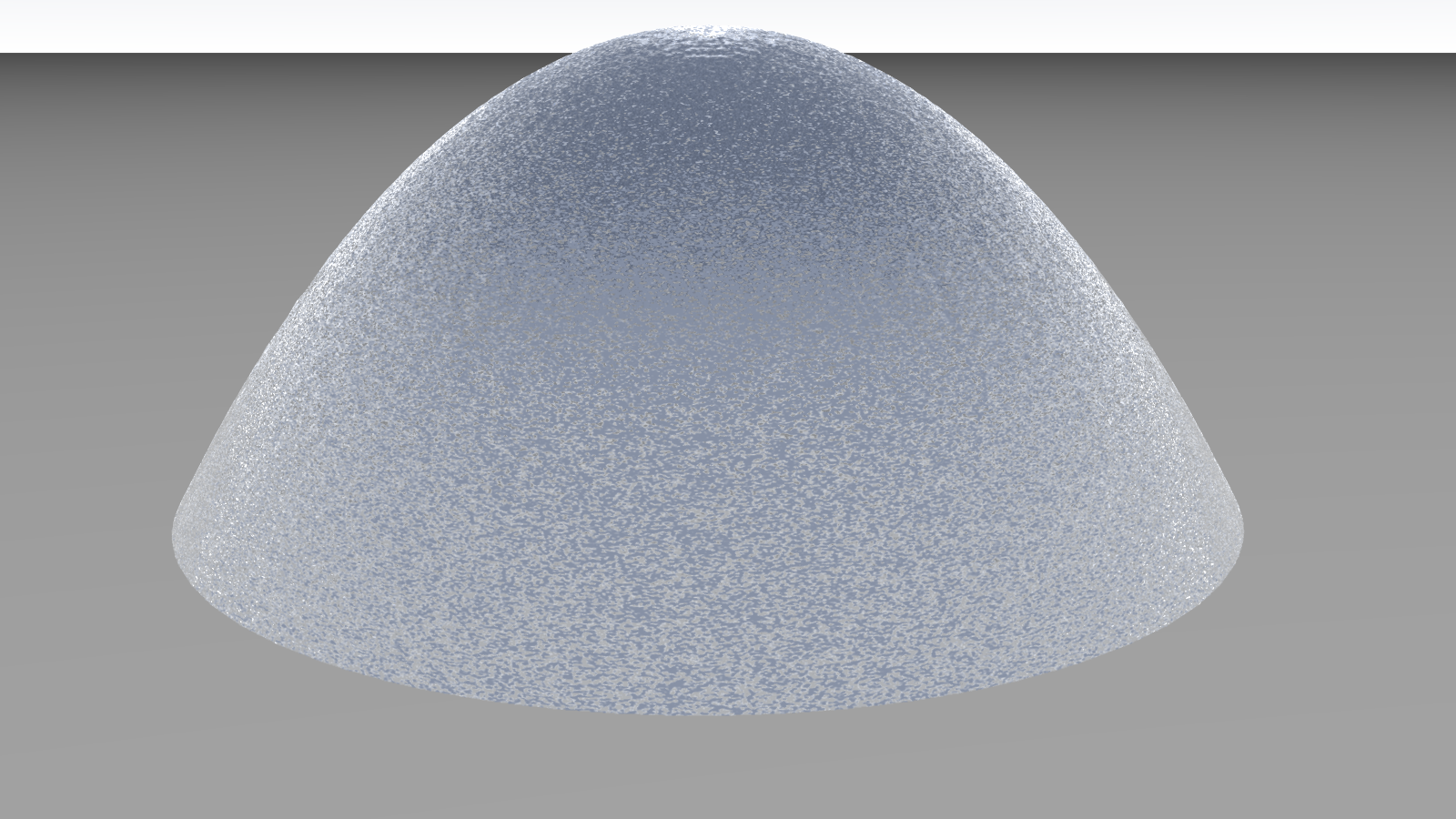}
        \subcaption{Initial droplet surface with $\theta = \pi/6$.}
    \end{subfigure}
    ~
    \begin{subfigure}[t]{0.45\textwidth}
        \centering
        \includegraphics[width=1.0\textwidth]{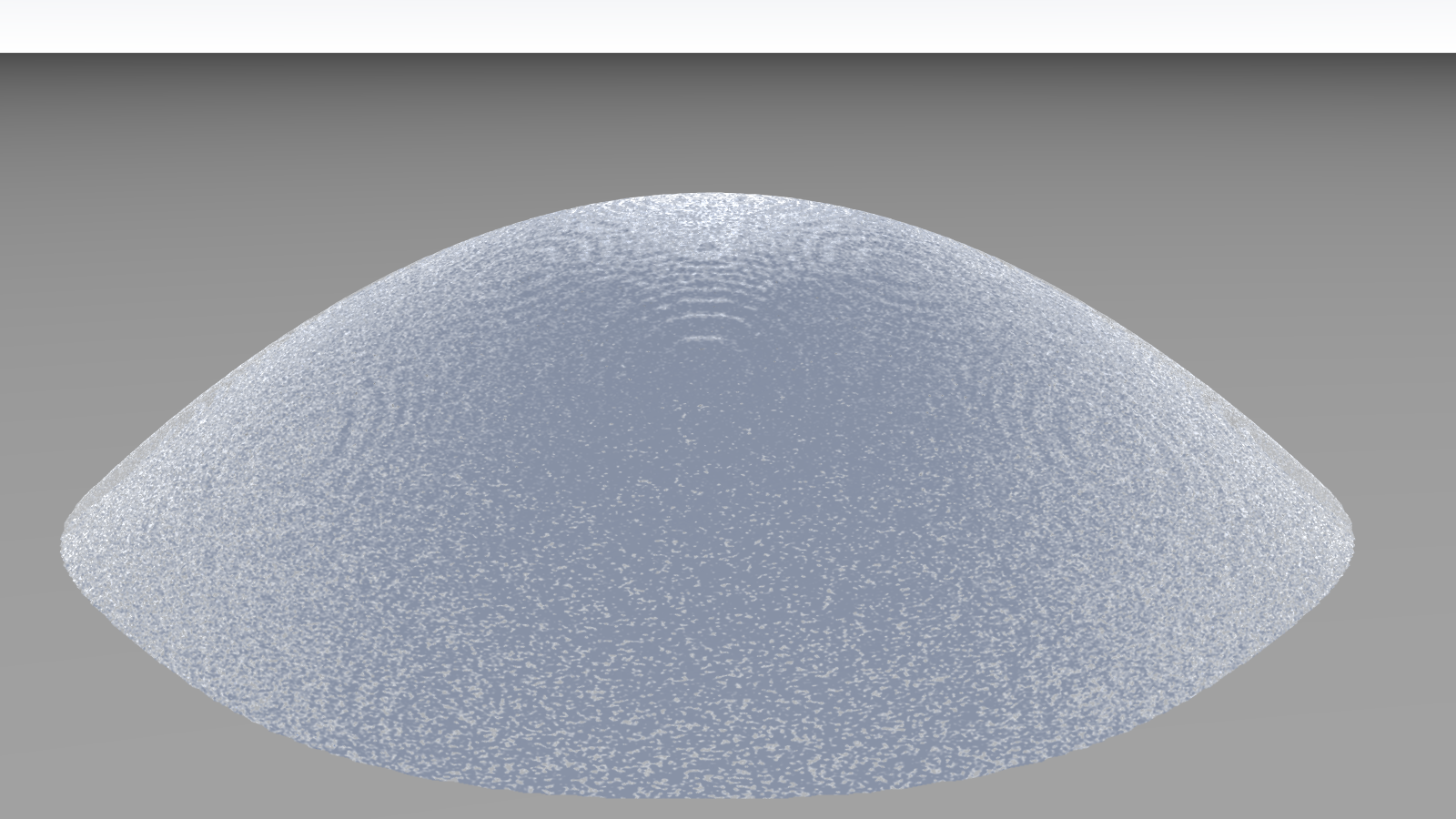}
        \subcaption{The droplet surface spreads and relaxes to the expected shape, as dictated by the equilibrium contact angle.}
    \end{subfigure}
    \caption{Relaxation of an out-of-equilibrium droplet.}
    
    \label{fig:drop_contour}
\end{figure*}

Let us consider the problem of a droplet, deposited on a smooth substrate
with an apparent contact angle $\theta > \theta_{eq}$, which spreads to relax
to a shape dictated by the equilibrium contact angle $\theta_{eq}$.
The equilibrium contact angle quantifies the wettability of a given substrate by a 
certain liquid and can be calculated using Young's equation \cite{RevModPhys.57.827}
\begin{equation}\label{eq:Young}
    \gamma \cos\theta_{eq} = \gamma_{SL} - \gamma_{SG} ,
\end{equation}
with $\gamma_{SL}$ and $\gamma_{SG}$ being the surface tensions between solid/liquid and solid/gas, respectively. 

In our simulations we set the equilibrium contact angle through the disjoining 
pressure (Eq.~(\ref{eq:disjoiningP})).
{
In order to comply as much as possible with the thin film assumptions, we limit ourselves to relatively small
contact angles.}
%Since our approach effectively reduces the 
%dimensionality of the problem from three
%to two, we are not able to impose contact angles larger then $\pi/2$.
%Otherwise, the height field would be a multivalued function, which violates the
%mathematical model.
%Strictly speaking, to satisfy the lubrication approach we are limited to small angles. 
%In recent references it is demonstrated, however, that although beyond the limits of lubrication theory, e.g. %contact angles of up to $5\pi/6$,
%spreading still satisfies the Cox-Voinov law, which is derived from lubrication theory~\cite{RevModPhys.81.739, %doi:10.1146/annurev-fluid-011212-140734, doi:10.1063/1.2171190}.

To probe the spreading, on a $512\times 512$ lattice we initialize a droplet, whose
surface is given by the expression
\begin{equation}
    h(x,y,0) = \sqrt{R^2 - (x-x_0)^2 - (y-y_0)^2} - R\cos\theta,
\end{equation}
with $R\sin\theta \approx 100\Delta x$  ($\theta>\theta_{eq}$) being the radius of the droplet with a spherical cap shape, and $(x_0,y_0)$ its center. The droplet is placed in the center of the lattice, i.e. $x_0 = y_0 = 256\Delta x$.  

In Fig.~\ref{fig:drop_contour} we show such an initial shape, with contact angle $\theta = \pi/6$, and the equilibrium shape with contact angle $\theta_{eq} = \pi/12$. 

To extract the contact angle from our data we impose that the shape at all times is close to the shape of a spherical cap, such that we are able to calculate the contact angle at any time using the initial angle and radius to obtain the volume
\begin{equation}
\label{eq:spherical_cap_vol}
    V = \frac{\pi}{3}R^3(2+\cos{\theta})(1-\cos{\theta})^2 .
\end{equation}
Since our method is mass conserving, the volume of the droplet is by construction conserved. Measuring both the height of the droplet $h_d(t)$ and the diameter of the spherical cap $r(t)/2$ we are able to recalculate the time dependent sphere radius as
\begin{equation}
    R(t) = \frac{r(t)^2+ h_d(t)^2}{2h_d(t)} 
\end{equation}
and can solve Eq.(\ref{eq:spherical_cap_vol}) again for the contact angle $\theta(t)$. We cross-checked our results with an alternative approach to calculate the angle given by
\begin{equation}
    \theta(t) = \sin{\left(\frac{r(t)}{R(t)}\right)}^{-1}.
\end{equation}

Let us stress that the shape is indeed very close to a spherical cap. As mentioned in 
Section \ref{sec:method}, in fact, a sufficiently accurate finite difference scheme
is required, as the one in Eqs. (\ref{eq:derivative}-\ref{eq:n_laplace})~\cite{THAMPI20131}. 
In particular, we note that the isotropy of the pressure gradient is of utmost importance: a simple scheme with two-point centered derivatives~\cite{zhou2004lattice} yields squared equilibrium droplet shapes.
 
\begin{figure}
    \centering %Figure without inset is also available, however I personally like this more.
    \includegraphics[width=0.48\textwidth]{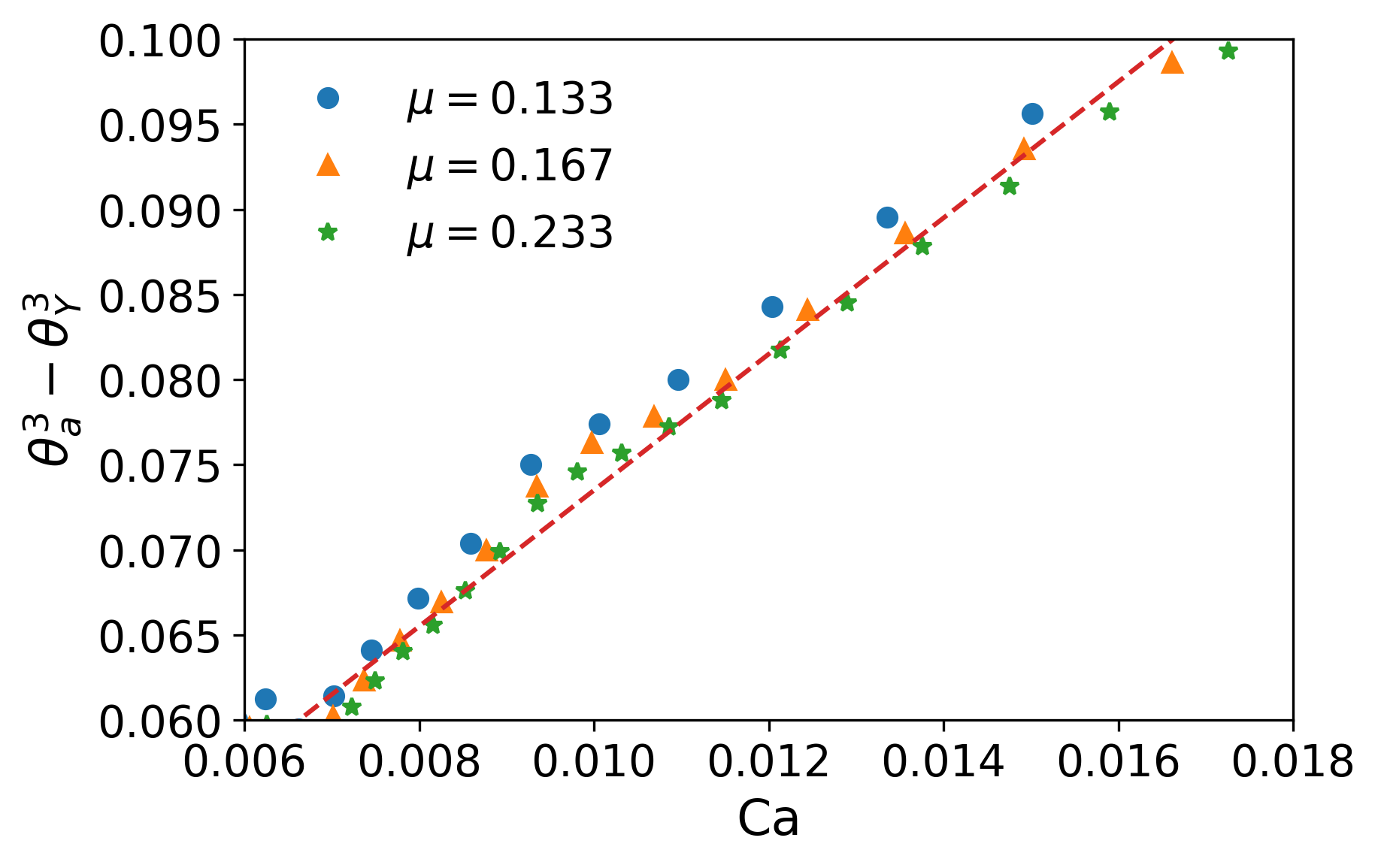}
    \caption{(Color online) Difference of cubed instantaneous and equilibrium contact angles, $\theta_{num}^3-\theta_{eq}^3$, vs. capillary number $Ca$ for a spreading droplet; the dashed line shows a linear dependence (consistent with the Cox-Voinov law). The different symbols represent different viscosities, while the dashed line is a linear function of the capillary number.
    \label{fig:Cox-Voinov}}
\end{figure}
The spreading dynamics can be investigated even more quantitatively in terms of the 
so-called Cox-Voinov law and Tanner's law~\cite{Tanner_1979}. The first one relates the apparent contact angle to 
the velocity $U$ of the spreading front (the contact line), at various times, by 
$\theta^3 - \theta_{eq}^3 \propto Ca$. The capillary number $Ca$ is defined
as $Ca=\mu U/\gamma$~\cite{doi:10.1146/annurev-fluid-011212-140734}.
In Fig.~\ref{fig:Cox-Voinov} we plot $\theta^3(t) - \theta_{eq}^3$ vs $Ca(t)$ from a numerical simulation of a spreading drop: a good linear scaling, in agreement with the Cox-Voinov law, is 
observed, as highlighted by the dashed line. 

\begin{figure}
    \centering
    \includegraphics[width=0.48\textwidth]{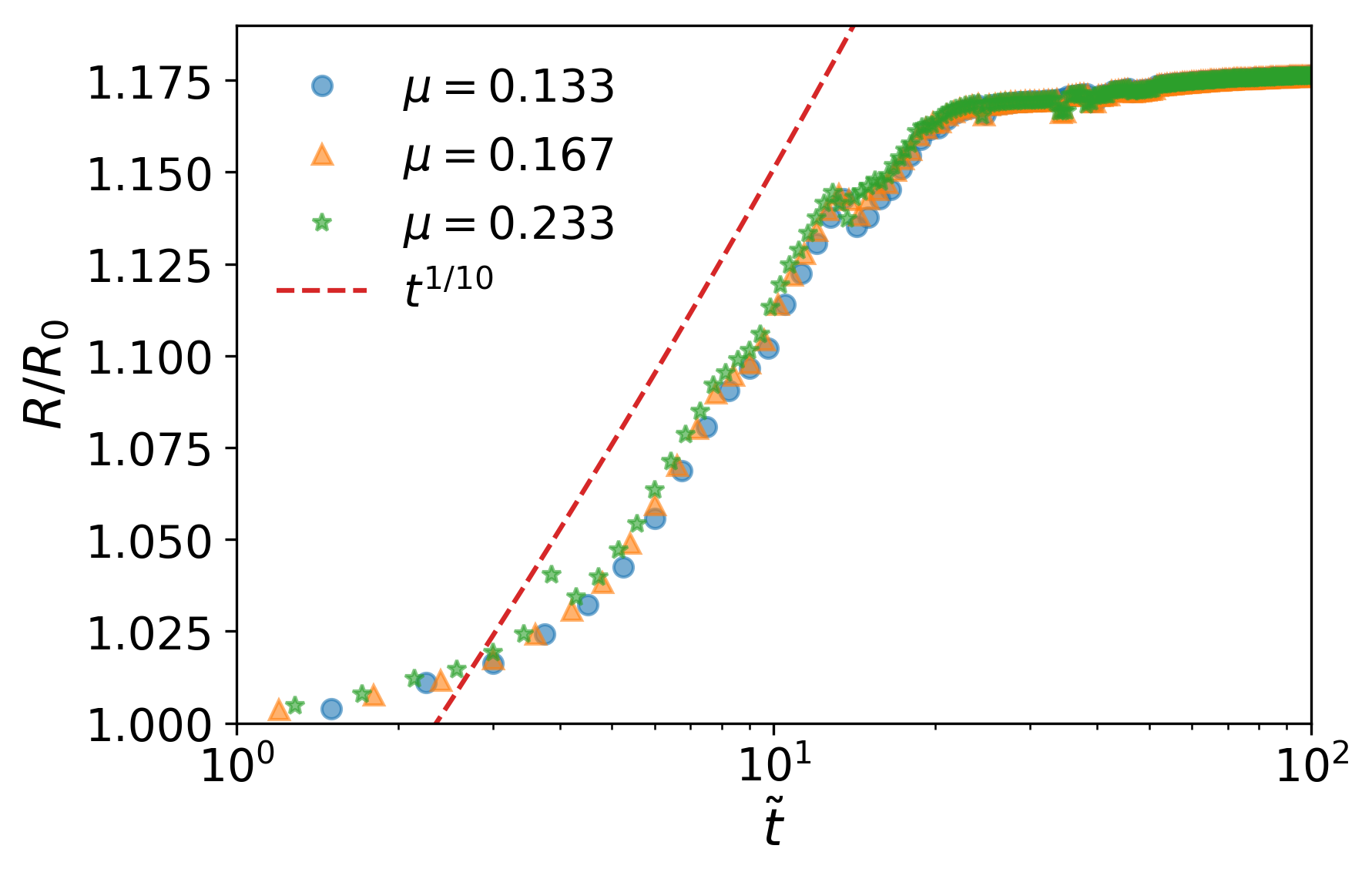}
    \caption{(Color online) Time evolution of the droplet base radius of a spreading droplet; the dashed red line shows a $\tilde{t}^{1/10}$ power law (In consistence with Tanner's law). As in Fig.~\ref{fig:Cox-Voinov} different symbols refer to different viscosities. The radius clearly grows with the predicted power law until it saturates. Upon rescaling the time with $\tau_{\mbox{\tiny{cap}}}$ the curves of all three viscosities collapse into a single one.}
    \label{fig:Tanners_law}
\end{figure}
The Tanner's law which states that the radius of the droplet grows with time as
\begin{equation}
\label{eq:tanners_law}
    R(t)\approx \left[\frac{10\gamma}{9B\mu}\left(\frac{4V}{\pi}\right)^3 t\right]^{1/10},
\end{equation}
with the constant $B$ being such that 
$B^{1/10} \approx 1.2$.  
%The important point here is that the 
%exponent of $t$ in Eq.~(\ref{eq:tanners_law}) can be %measured in experiments. It turns 
%out that $n=1/10$ is a good fit \cite{rioboo2002time, jambon-puillet_carrier_shahidzadeh_brutin_eggers_bonn_2018, cazabat1986dynamics, CHEN198860}. 
In Fig.~\ref{fig:Tanners_law} 
we plot the measured radius of the droplet divided by its initial radius $R_0$ as a function of the dimensionless time $\tilde{t} = t/\tau_{\mbox{\tiny{cap}}}$ 
(here $\tau_{\mbox{\tiny{cap}}} = \frac{\mu R}{\gamma}$). For the three viscosities considered 
in Figs.~\ref{fig:Cox-Voinov},\ref{fig:Tanners_law} our capillary times are 
$\tau_{\mbox{\tiny{cap}}} = [1333, 1667, 2333] \Delta t$. 
We see a saturation at $R/R_0=1.17$ because the droplet is very close to 
its equilibrium shape. During the growth phase the radius follows indeed a power law 
in $\tilde{t}$ with exponent $1/10$, which is shown by the red dashed line, in agreement with Tanner's prediction and 
experimental results \cite{rioboo2002time, jambon-puillet_carrier_shahidzadeh_brutin_eggers_bonn_2018, cazabat1986dynamics, CHEN198860}.
We further notice that within our
simulations the droplet needs about $12\tau_{\mbox{\tiny{cap}}}$ to reach its equilibrium shape.

% Lastly, we discuss problems involving higher contact angles and flow profiles. 
% Since our approach effectively reduces the dimensionality of the problem from three
% to two, we are not able to impose contact angles larger then $\pi/2$. 
% Strictly speaking, to satisfy the lubrication approach we are limited to small angles. 
% In recent references~\cite{RevModPhys.81.739, doi:10.1146/annurev-fluid-011212-140734, doi:10.1063/1.2171190} it is demonstrated, however, that although beyond the limits of lubrication theory, e.g. contact angles of up to $5\pi/6$,
% spreading still satisfies the Cox-Voinov law, which is derived from lubrication theory. 
% Assuming that a contact angle is larger than $\pi/2$ we observe unstable behaviour as
% the height field is no longer be a well defined function. 

\subsection{A Sliding droplet}
\begin{figure}
    \centering
    \includegraphics[width=0.45\textwidth]{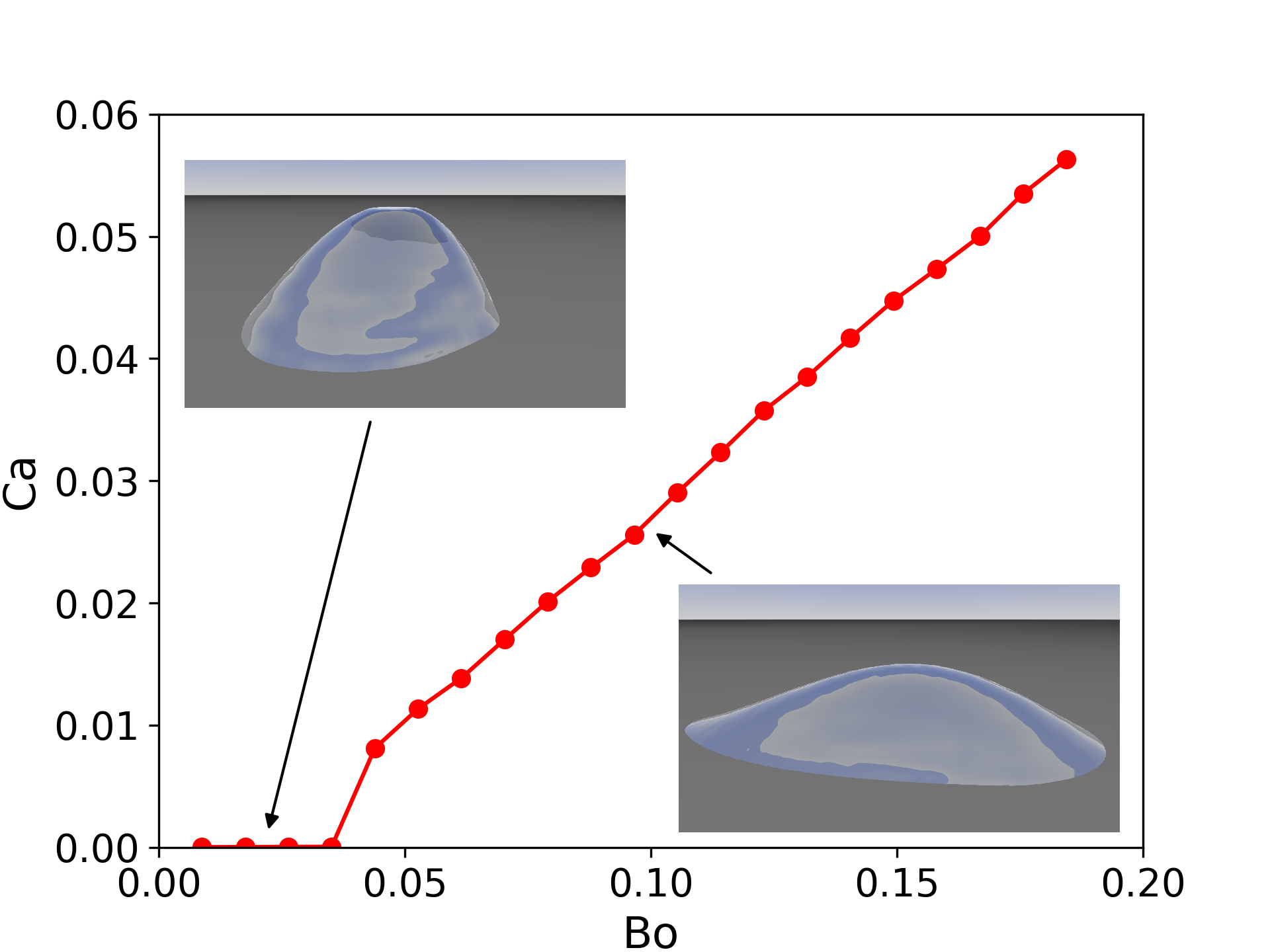}
    \caption{(Color online) $Ca$ vs $Bo$ for a sliding droplet: notice that a finite minimum forcing (corresponding to $Bo_c$) is needed to actuate the droplet. For $Bo > Bo_c$ a linear relation, $Ca \sim Bo$, is observed. In the insets we show the shape of the droplet for both, the pinned (upper left) as well as the sliding (lower right) case.
    }
    \label{fig:CaBo}
\end{figure}

As a further validation case we consider the sliding of a droplet on an inclined plane.
For a droplet to slide over an inclined plane, a minimum tilting angle $\alpha >0$ of the substrate is required~\cite{Furmidge}, which in our case is due to the friction term Eq.~(\ref{eq:alphafric}). Until this critical angle is reached energy is stored in the deformation of the surface as the upper left inset in Fig.~\ref{fig:CaBo} shows. Above such a critical angle, a linear relation between the terminal sliding velocity $U_{\infty}$ and the gravitational force $\propto m g \sin \alpha$ is observed~\cite{Podgorski,Kim,Sbragaglia}; in dimensionless numbers such 
behaviour is expressed by 
\begin{equation}\label{eq:CaBo}
Ca \propto Bo - Bo_c,
\end{equation}
where the capillary number is based on $U_{\infty}$ and $Bo$ is the so called Bond number, given by
\begin{equation}
  Bo = (3V/4\pi)^{2/3}\rho g \frac{\sin\alpha}{\gamma}.
\end{equation}
$Bo_c$ is the critical Bond number, defined in terms of the critical 
tilting angle $\alpha_c$.
In Fig.~\ref{fig:CaBo} we plot $Ca$ vs $Bo$ from our numerical simulations, showing that the phenomenology described by Eq.~(\ref{eq:CaBo}) is 
indeed reproduced, i.e. the onset of sliding takes place at a finite 
forcing, beyond which the linear scaling $Ca \sim Bo$ is fulfilled. 

\begin{figure*}
  \begin{subfigure}[t]{0.32\textwidth}
    \centering\includegraphics[width=1.0\textwidth]{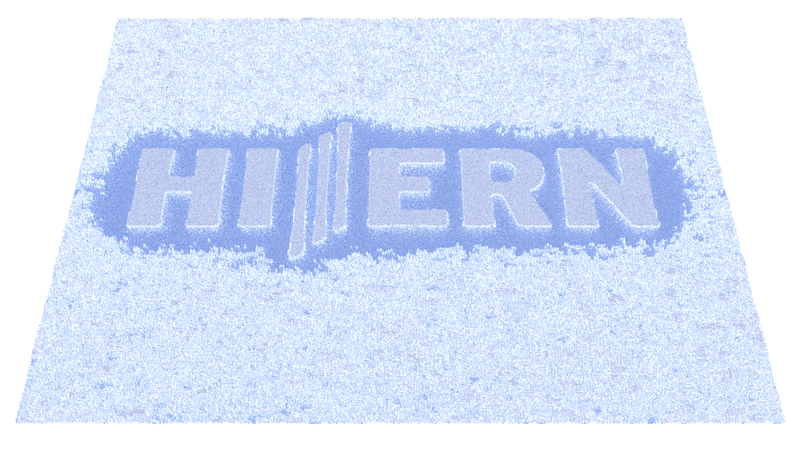}
    \caption{t $=2400\Delta$t}
  \end{subfigure}
  \begin{subfigure}[t]{0.32\textwidth}
    \centering\includegraphics[width=1.0\textwidth]{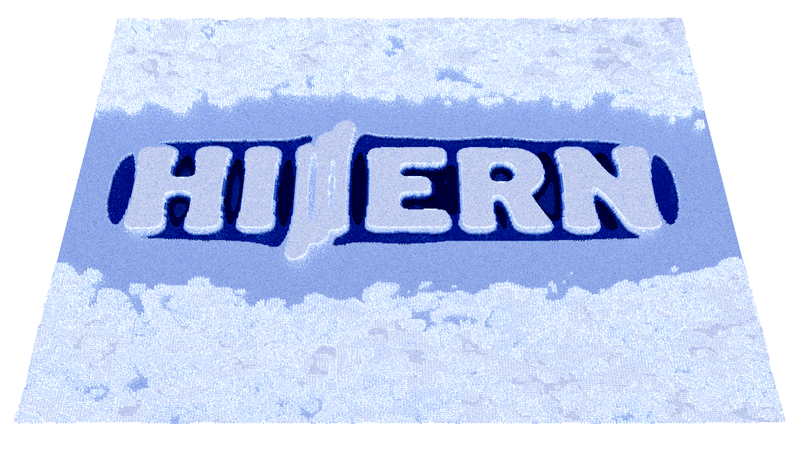}
    \caption{t $=16800\Delta$t}
  \end{subfigure}
  \begin{subfigure}[t]{0.32\textwidth}
    \centering\includegraphics[width=1.0\textwidth]{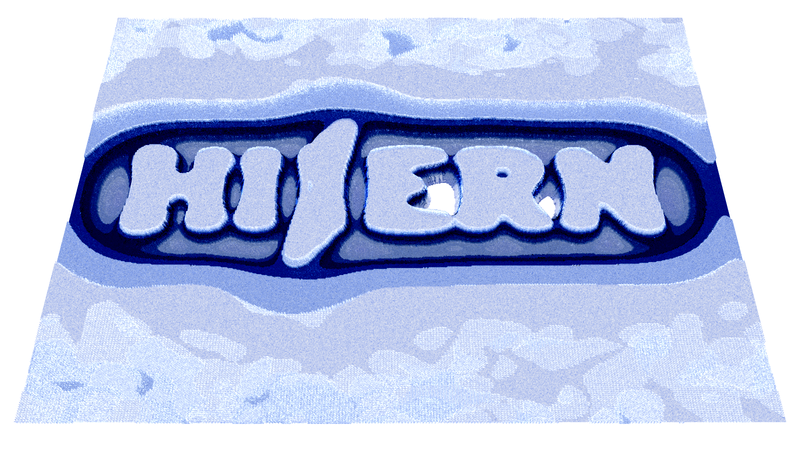}
    \caption{t $=97200\Delta$t}
  \end{subfigure}
  \caption{(Color online) Time evolution of the free surface on a chemically patterned substrate on a 512x512 $\Delta x^2$ domain. Varying the contact angle between the letters and the rest of the substrate yields the shown dewetting pattern. The letters are more wettable then the rest. To emphasis the process we use a color gradient raging from dark blue to light blue.
  Starting from a randomly perturbed film height, the fluid starts to dewet the pattern (a) and after 2400$\Delta t$ the letters and a surrounding rim structure are clearly visible. Towards the end of the simulation (c), the instability of the thin film also leads to film rupture. Holes form between the letters E, R and N.  
  }
  \label{fig:Logo_evolution}
\end{figure*}

\subsection{Dewetting of liquid films}
In order to show-case the capabilities of our method in handling more complex physics scenarios, we finally consider the dewetting of a chemically patterned substrate~\cite{Kargupta,Brasjen}. 
This is easily made possible within the code by introducing a space-varying equilibrium contact angle, $\theta_{eq}(x,y)$, in Eq.~(\ref{eq:disjoiningP}); in this way we can tune the local wettability of the substrate. 
Fig.~\ref{fig:Logo_evolution} shows a liquid film which is initialized with thickness $h(x,y,0)$ randomly fluctuating in space around its mean value $h_0$, by a small percentage ($\approx 0.01\%$)  of it (panel (a)). A partially wettable substrate is patterned in such a way that the contact angle is lower on a region defining a logo. The total domain contains 512x512 lattice nodes. With this domain size a letter contains around 130 lattice nodes in y-direction and about 60 lattice nodes in x-direction.Using the initial height $h_0$ of the film as characteristic length scale we get a capillary time of $\tau_{\mbox{\tiny{cap}}} \approx 20\Delta t$. As the film dewets, liquid moves toward the letters of the logo, the surrounding film becomes thinner and eventually the logo becomes visible.

\section{Computational aspects}
We use \textit{OpenACC} directives to allow our code to run on accelerator devices, such as 
Graphics Processing Units (GPUs), while being able, at the same time, to exploit the well known good scaling properties of the LBM on parallel machines~\cite{Chandrasekaran:2017:OPC:3175812}. 
\textit{OpenACC} is particularly versatile in terms of programmability since it only requires a few lines of code to allow us to harness the power of state of the art accelerators. In this sense \textit{OpenACC} is very similar to \textit{OpenMP} {and} more readable as well as much easier to program than \textit{CUDA}.
%{\textit{OpenACC} helps us to port the code to the GPU with two main features. First being the data region, as an 
%array can be generated and manipulated on the device memory, which is shared among all threads. Second being the compute regions. %Where the programmer simply adds a \#pragma statement in front of an parallelizable loop.}

The performance of a LBM code is commonly measured in Million Lattice Updates Per Second (\textbf{MLUPS}), defined as
\begin{equation}
    \textbf{MLUPS}=\frac{A\times n}{t_{sim}\times 10^6},
\end{equation}
with $A=L_x\times L_y$ being the area of the lattice,
where $L_x,L_y$ are the number of lattice nodes in $x$ and $y$ directions. The number of iterations is given by $n$. The time needed to compute the $n$ interations is called $t_{sim}$ (in seconds).
In Tab.~\ref{tab:efficiency} we provide benchmark data comparing the performance of a Nvidia GTX 1080TI, a Nvidia Quadro K2200 and a single core of an Intel i7-4790 @ 3.6GHz CPU.
Due to the limited amount of memory available on the Quadro K2200, it is not possible to run a simulation of size $4096^2$ on this card. Such a simulation requires about 4.8 GB local memory, while the Quadro K2200 only supplies 4 GB. In particular the speedup gained by using a GTX 1080TI is outstanding and corresponds to about 24-92 times the performance of a single core of the Intel CPU. Assuming perfect scaling on the CPU and using all 4 physical cores, the simulation on the GPU would be faster by a factor between 6 and 23. The speedup depends on the size of the lattice and in order to keep the pipelines on the GPU filled, a minimum loop size is needed. In addition, data transfer between \textit{host} and \textit{device} is a known bottleneck impacting the performance of GPU based simulations. This is obviously also the case for our code -- even though such data transfer is only needed when files are written to disk.
 
\begin{table}
 \begin{tabular}{|c|c c c c c c |}
 \hline
  Lattice/Accelerator & $128^2$ & $256^2$ & $512^2$ & $1024^2$ & $2048^2$ & $4096^2$ \\ \hline
  GTX 1080TI & $157.6$ & $279.2$ & $382.6$ & $414.9$ & $404.7$ & $395.6$ \\ \hline
  Quadro K2200 & $33.5$ & $42.9$ & $46.6$ & $48.2$ & $49.0$ & $X$ \\ \hline
  i7-4790 & $6.4$ & $5.8$ & $4.5$ & $4.6$ & $4.5$ & $4.3$ \\
  \hline
 \end{tabular}
 \caption{Performance analysis based on a MLUPS measurement. The different columns relate to different lattice sizes, while the rows correspond to the two GPUs and one CPU used. All simulations are run for 100000$\Delta t$ with FP64 double precision.}
 \label{tab:efficiency}
 \end{table}

\section{Conclusions}\label{sec:conclusions}
We have presented a novel lattice Boltzmann model for the numerical simulation of thin liquid film 
hydrodynamics, featuring explicitly relevant properties of interface physics, such as surface tension and disjoining pressure.

We validated our method against a relevant test case, namely the Rayleigh-Taylor instability, where the critical wavenumber as well as the growth and damping of wavemodes are correctly reproduced. Our simulations of droplets on substrates showed that droplets initiated out of equilibrium attain their equilibrium contact angle and that our method correctly reproduces the Cox-Voinov law as well as Tanner's law. Furthermore, our approach allows to simulate the dynamics of sliding droplets and even complex dewetting scenarios.

Our \textit{OpenACC} enabled simulation code allows for a massive improvement of the performance: with modern GPU cards at hand simulations using large lattice sizes and requiring many time steps can be run on a single workstation without the need for access to high performance computing resources.

In the future, we plan to extend our work towards systems which could hardly be tackled by traditional methods: from the dynamics of individual droplets on complex shaped substrates we plan to move to large numbers of droplets in order to understand the statistical properties of collective droplet motion on chemically structured substrates. Finally, a possible application of our method could be the simulation of full lab-on-chip devices with highly resolved channels, junctions, etc..

\begin{acknowledgements}
The authors acknowledge financial support by the Deutsche 
Forschungsgemeinschaft (DFG) within the Cluster of Excellence ``Engineering of Advanced Materials'' (project EXC 315) (Bridge Funding). The work has been partly performed under the Project HPC-EUROPA3 (INFRAIA-2016-1-730897), with the support of the EC Research Innovation Action under the H2020 Programme; in particular, S. Z. gratefully acknowledges the support of Consiglio Nazionale delle Ricerche (CNR) and the computer resources and technical support provided by CINECA.
\end{acknowledgements}

\appendix
\section{}
\label{app:only}
\begin{figure*}

  \begin{subfigure}[t]{0.32\textwidth}
    \centering\includegraphics[width=1.0\textwidth]{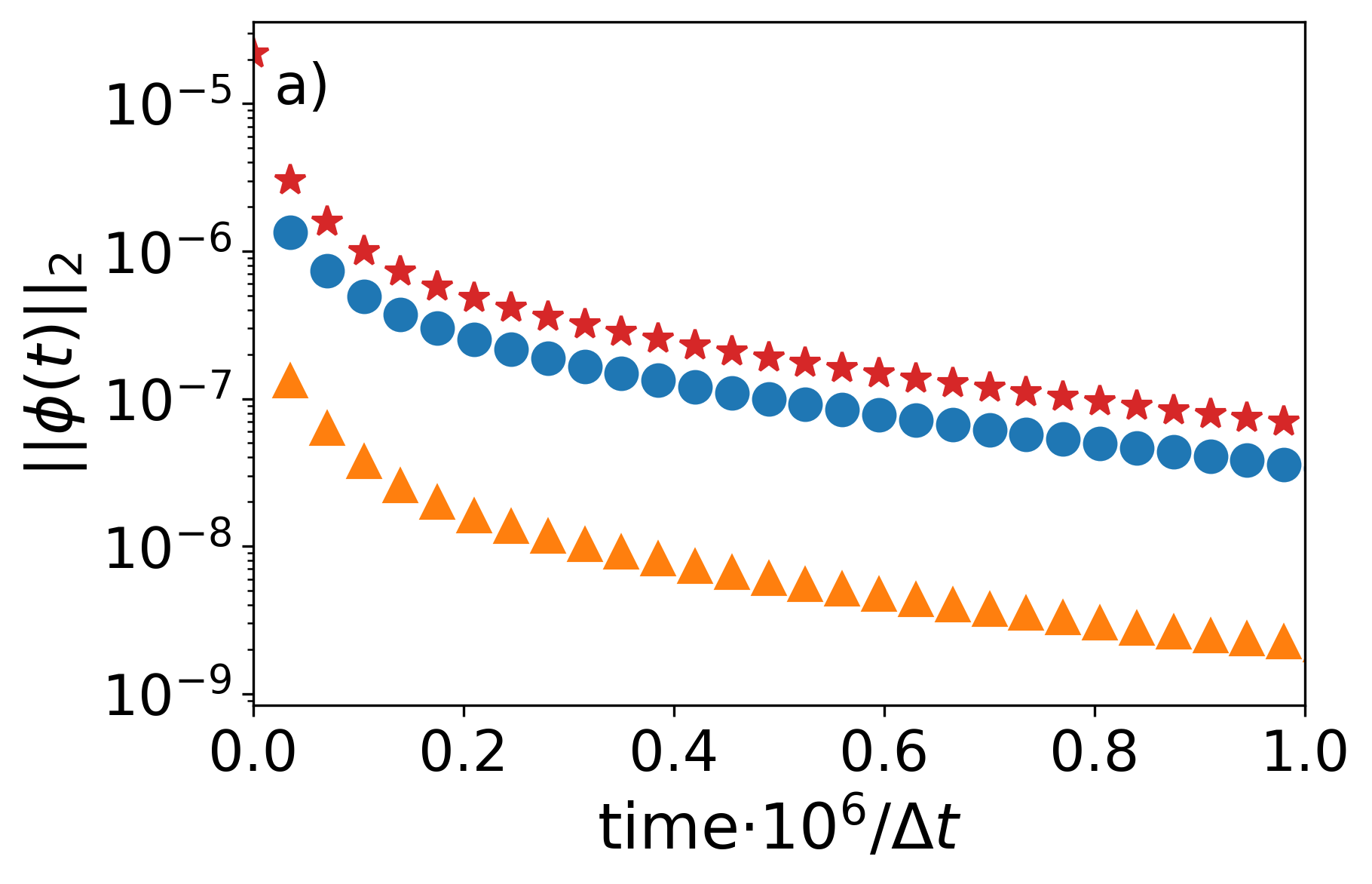}
  \end{subfigure}
  \begin{subfigure}[t]{0.32\textwidth}
    \centering\includegraphics[width=1.0\textwidth]{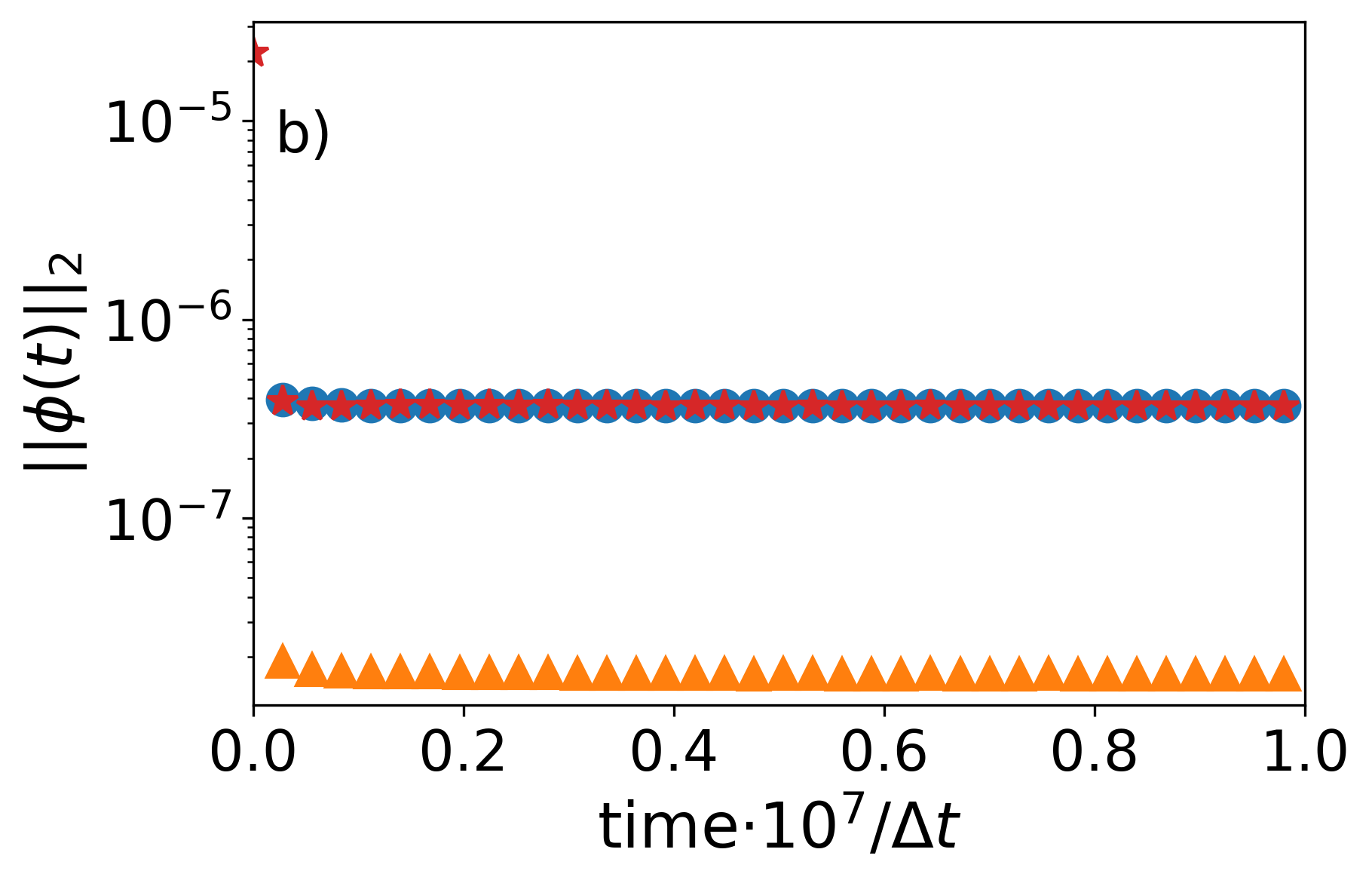}
  \end{subfigure}
  \begin{subfigure}[t]{0.32\textwidth}
    \centering\includegraphics[width=1.0\textwidth]{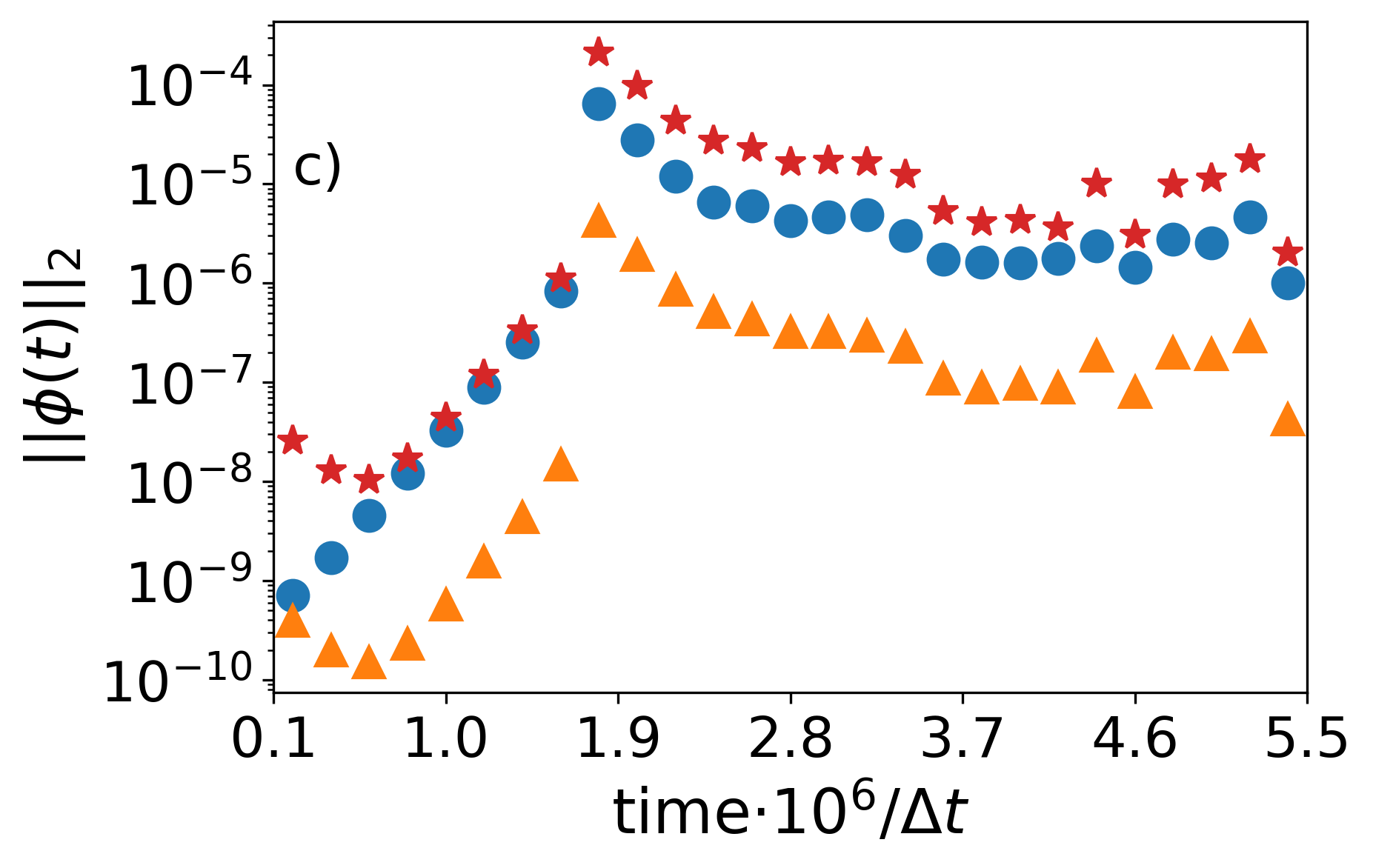}
  \end{subfigure}
  \caption{(Color online) Time evolution of the 
  {$\mathbf{L}_2$ norm} (as defined by equation 
  (\ref{eq:magnitude})) of the $x$-component of the terms appearing on the right hand side of the second of equations (\ref{eq:hydro2}). The plot is in log-lin scale.
  The panels refer to three different numerical experiments: (a) spreading droplet, (b) sliding droplet and (c) thin film dewetting. The symbols correspond to: 
  film pressure gradient, $-h\partial_x p_{\mbox{\tiny{film}}}$, (\textcolor{pyred}{$\star$}); friction, $-\nu \alpha(h) u_x$, (\textcolor{pyblue}{$\bullet$}); longitudinal dissipation terms 
  $\nu \nabla^2 (h u_x) + 2\nu \partial_x (\nabla \cdot (h\mathbf{u}))$, (\textcolor{pyorange}{$\blacktriangle$}).}
  \label{fig:Referee_1}
\end{figure*}

As anticipated above, we provide here a numerical validation of the assumptions on
effectively negligible terms that lead from Eq.~(\ref{eq:hydro}) to Eq.~(\ref{eq:hydro3}). 
To this aim, we report in Fig.~\ref{fig:Referee_1}, for each of the term under scrutiny, 
the time evolution of a {$\mathbf{L}_2$-norm}, defined for a generic scalar field $\phi(\mathbf{x},t)$ as 
{
\begin{equation}\label{eq:magnitude}
||\phi(t)||_2 = \left(\frac{1}{N^2}\sum_{i=1}^{N}\sum_{j=1}^{N}\left(\phi(x_i,y_j,t)\right)^2\right)^{\frac{1}{2}},
\end{equation}
}
where the double sum is extended to the whole two-dimensional domain. Three case-studies are analyzed
(corresponding to the three panels in Fig.~\ref{fig:Referee_1}), namely
(a) a sessile droplet spreading on a substrate with an equilibrium contact angle smaller than the initial one, 
(b) a droplet sliding under the action of a body force and (c) the dewetting of a substrate. 
We compare, for each simulation, the {$||\phi(t)||_2$} for the $x$-component\footnote{Similar results are found also 
for the $y$-component.} of the gradient of the film pressure, $-h\partial_x p_{\mbox{\tiny{film}}}$, 
of the friction term, $-\nu \alpha(h) u_x$, and of the longitudinal viscous terms, 
$\nu \nabla^2 (h u_x) + 2\nu \partial_x (\nabla \cdot (h\mathbf{u}))$
{(the advection term, $\nabla \cdot (h \mathbf{u}u_x)$ 
is for all cases orders of magnitude smaller than the other terms, therefore we decided to omit it from the comparisons 
in figure Fig.~\ref{fig:Referee_1}).}
%The correctness of the assumptions is particularly
%evident in the {sliding} droplet numerical experiment (panel ({b})), where the time %evolution 
%of the {$\mathbf{L}_2$ norm} suggests that the friction term perfectly balances the film pressure %gradient, which amounts to effectively 
%solving the lubrication equation (as {discussed} in Sec.~\ref{sec:method}). 
{We observe
that the gradient of the film pressure and the friction are dominant, with the $\mathbf{L}_2$ norm of the longitudinal dissipation term being always, roughly, less than $10\%$ of the friction contribution.}

%\bibliography{Ref}

\end{document}